\documentclass[12pt]{article}
\usepackage[english,activeacute]{babel}
\usepackage{natbib}
\usepackage{comment}
\usepackage{float}
\usepackage[hidelinks]{hyperref}
\usepackage{mathrsfs}
\usepackage{enumitem}
\usepackage[font={small,it}]{caption}
\usepackage{amsmath,amsfonts,amsthm,amssymb}
\usepackage{bm,rotating,multirow,dsfont,graphicx}
\usepackage[usenames, dvipsnames]{color}
\usepackage{url}
\usepackage{multicol}
\usepackage{multirow}
\usepackage[T1]{fontenc}
\usepackage{flafter}
\usepackage{appendix}
\usepackage{subfigure}
\usepackage{xcolor}
\usepackage{soul}
\usepackage{setspace}
\makeatletter
\def\hlinewd#1{%
	\noalign{\ifnum0=`}\fi\hrule \@height #1 %
	\futurelet\reserved@a\@xhline}
\makeatother
\def\spacingset#1{\renewcommand{\baselinestretch}{#1}\small\normalsize}\spacingset{1}
\def\@roman#1{\romannumeral #1}

\begin{document}

\title{Bayesian Sociality Models: A Scalable and Flexible Alternative for Network Analysis}

\date{}

\author{
    Juan Sosa\footnote{Corresponding author: jcsosam@unal.edu.co.}, Universidad Nacional de Colombia, Colombia \\
    Carlos Martínez, Universidad Nacional de Colombia, Colombia    
}

\maketitle


\begin{abstract}
Bayesian sociality models provide a scalable and flexible alternative for network analysis, capturing degree heterogeneity through actor-specific parameters while mitigating the identifiability challenges of latent space models. This paper develops a comprehensive Bayesian inference framework, leveraging Markov chain Monte Carlo and variational inference to assess their efficiency-accuracy trade-offs. Through empirical and simulation studies, we demonstrate the model’s robustness in goodness-of-fit, predictive performance, clustering, and other key network analysis tasks. The Bayesian paradigm further enhances uncertainty quantification and interpretability, positioning sociality models as a powerful and generalizable tool for modern network science.
\end{abstract}


\spacingset{1.1} 

\section{Introduction}

The study of social networks is crucial for understanding the structural and dynamic properties of complex systems where interactions among entities define collective behaviors. Networks appear in numerous disciplines, from sociology and epidemiology to computer science and biology, serving as fundamental representations of relationships among individuals, organizations, or objects. Statistical modeling of networks provides valuable tools for uncovering latent structures, predicting missing links, and understanding the generative processes that shape these complex systems \citep{barabasi2016, newman2018, kolaczyk2020statistical, menczer2020}.

Bayesian modeling is a powerful framework for network analysis, offering a probabilistic approach that integrates external information, accounts for uncertainty, and enables robust inference (ensuring stability, reliability, and improved generalization, particularly in uncertain, sparse, or structured data settings). Its flexibility allows for application across various network models, including random graph models, exponential random graph models (ERGMs), stochastic block models (SBMs), and latent space models (see \citealt{sosa2021review} for a review). Each of these models captures different structural aspects of networks. ERGMs, for instance, emphasize endogenous dependencies such as reciprocity and transitivity \citep{snijders2002, robins2007}, while SBMs are particularly effective for identifying community structures by clustering nodes into latent groups \citep{nowicki-2001, airoldi2008}. Latent space distance models, in contrast, embed nodes in a continuous space where their relative distances define tie probabilities, making them well-suited for modeling homophily and transitivity \citep{hoff2002, schweinberger2003, sosa2021review}.

A key strength of Bayesian inference in network modeling lies in its ability to incorporate hierarchical structures and hyperparameters, allowing for a more detailed representation of node-level attributes and global network patterns \citep{handcock2007, durante2014}. Additionally, Bayesian methods support posterior predictive checks and model selection criteria such as the Deviance Information Criterion (DIC) and the Widely Applicable Information Criterion (WAIC), which are particularly effective for comparing competing network models \citep{gelman2014bayesian}. Recent advances in Bayesian computation, including Hamiltonian Monte Carlo and variational inference, have significantly improved the scalability and practicality of Bayesian network models, enabling their application to large-scale networks \citep{raftery2012, salter2013, kim2018}. Unlike frequentist approaches, which often struggle with intricate dependency structures and rely on asymptotic approximations, Bayesian inference provides posterior distributions, leading to more reliable uncertainty quantification, data-driven regularization, and interpretability.

Among Bayesian approaches to network modeling, the sociality model has received relatively little attention despite its strong potential for capturing heterogeneity in node connectivity. \citet{krivitsky2009} introduced a latent cluster random effects model incorporating sociality parameters to account for degree heterogeneity, yet this approach remains underexplored. Unlike latent space models, which primarily emphasize dyadic relationships, the sociality model explicitly integrates individual-level parameters that govern tie formation, allowing it to better capture variations in node popularity and centrality that cannot be fully explained by spatial proximity or cluster membership. Given the critical role of degree heterogeneity in real-world networks, a more thorough investigation of the sociality model is both necessary and timely.

In a similar spirit, \cite{hoff2021additive} introduces an ANOVA decomposition to evaluate across-row and across-column heterogeneity in a data matrix, effectively modeling sociality by capturing correlations within rows of the sociomatrix. Heterogeneity arises as nodes acting as ``senders'' exhibit varying sociability, influencing the across-row variance of row means, while differences in ``popularity'' contribute to across-column variance. Extensions of this kind of models are discussed in \cite{gill2001statistical} and \cite{li2002unified}. While \cite{hoff2021additive} notes that ANOVA is easy to implement, it overlooks the dual role of nodes as both senders and receivers, implying that row and column effects should be correlated. We argue that despite its simplicity, this approach remains highly competitive compared to more structured models.

In addition, the sociality model serves as a robust foundation for broader generalization. Its framework can be naturally extended to incorporate hierarchical structures in multilayer networks, temporal dynamics in evolving networks, and contagion mechanisms for modeling diffusion processes. Multilayer network models enhance traditional approaches by incorporating multiple interaction types within a single system, an area of growing interest explored in \citet{kivela2014}, \citet{boccaletti2014}, \citet{de2016physics}, and \citet{sosa2022latent}. Dynamic network models extend static representations by introducing temporal evolution, enabling the study of structural changes over time, as examined in \citet{snijders2001statistical}, \citet{sewell2015latent}, and \citet{durante2014}. Likewise, models for diffusion processes, which investigate the spread of ideas, information, and diseases, require an accurate representation of individual variability in connectivity, with major contributions from \citet{pastor2001epidemic}, \citet{barrat2008dynamical}, and \citet{rogers2014diffusion}. By effectively capturing degree heterogeneity, the sociality model offers a robust and flexible foundation for integrating and advancing these sophisticated network modeling approaches while retaining computational feasibility.

A key advantage of the sociality model over other existing approaches is its balance between mathematical tractability, computational efficiency, and interpretative flexibility. Unlike latent space models, which require high-dimensional optimization and often face identifiability challenges when estimating node positions, the sociality model incorporates actor-specific parameters that are both more interpretable and computationally efficient. This makes it particularly well-suited for large-scale networks, where traditional latent space methods struggle with scalability. Additionally, once fitted, the sociality model facilitates straightforward derivations of network clustering, degree heterogeneity assessment, and individual-level predictions, making it a valuable tool for network-based classification and segmentation. Unlike stochastic block models, which rely on predefined or inferred group memberships, the sociality model provides a more continuous and nuanced representation of individual connectivity patterns, avoiding rigid categorical assumptions while maintaining modeling flexibility.

This paper presents a comprehensive examination of the sociality model, with a particular focus on its Bayesian estimation using Markov chain Monte Carlo (MCMC) and Variational Inference (VI). By comparing these methods, we analyze their computational efficiency and inference accuracy, highlighting the trade-offs between sampling-based techniques and optimization-based approximations. Additionally, we evaluate the performance of the sociality model relative to latent space models, emphasizing goodness-of-fit and predictive accuracy across multiple network datasets. The Lazega law firm dataset, a widely studied benchmark in network analysis, is used to demonstrate how the sociality model effectively captures variations in individual connectivity patterns. Furthermore, a simulation study is conducted to assess the ability of MCMC and VI to recover latent structures and to evaluate their computational feasibility in large-scale network settings.

By systematically examining the sociality model, this work establishes its value as a viable alternative to existing Bayesian network models. The findings enhance the broader understanding of network heterogeneity, providing insights into the trade-offs between different Bayesian inference approaches and showcasing the model’s practical applications in empirical research. The remainder of the paper is organized as follows. Section 2 introduces the formal mathematical specification of the sociality model. Section 3 outlines computational techniques for Bayesian inference, comparing MCMC and VI. Section 4 presents empirical analyses, including real-world datasets and simulations. Section 5 reports the results, and Section 6 provides a discussion that contextualizes the findings and outlines potential directions for future research.

\section{Model}\label{sec_model}

Consider an undirected binary network represented by the upper triangular portion of the \( n \times n \) adjacency matrix \( \mathbf{Y} = [y_{i,j}] \), where each entry \( y_{i,j} \in \{0,1\} \) denotes the presence (\( y_{i,j} = 1 \)) or absence (\( y_{i,j} = 0 \)) of a connection between individuals \( i \) and \( j \). The sociality model for static undirected binary networks specifies the probability of a connection between two nodes as:  
\[
y_{i,j} \mid \theta_{i,j} \overset{\text{ind}}{\sim} \textsf{Ber}(\theta_{i,j}), \quad \text{for } i < j,
\]  
where:  
\[
\theta_{i,j} = \Phi(\mu + \delta_i + \delta_j),
\]  
and \(\Phi(\cdot)\) represents the cumulative distribution function (CDF) of the standard Normal distribution. This CDF, referred to as the probit link function, ensures that the probabilities \(\theta_{i,j}\) are confined to the interval \((0, 1)\). Although alternative link functions are available, such as the logit link, the probit link function is particularly favorable since it facilitates the introduction of auxiliary random variables, which simplifies calculations in methods like Gibbs sampling.

Under this formulation, \(\mu\) represents the global connectivity effect, where higher values indicate a greater overall tendency for connections to form within the network. The parameters \(\delta_1,\ldots,\delta_n\), captures the sociability of node \(i\). Positive values of \(\delta_i\) indicate a higher individual propensity to form connections, while negative values suggest a reduced tendency. These \(\delta_i\) parameters account for individual differences in connectivity, allowing the model to capture patterns of centrality and isolation within the network. This parameterization allows the model to capture both global and individual variability in connectivity, providing a flexible framework for analyzing undirected binary networks.

To perform fully Bayesian inference on \(\mu\) and the \(\delta_i\), it is necessary to specify prior distributions for these parameters. Specifically, it is assumed that \(\mu \sim \textsf{N}(0, \sigma^2)\) and \(\delta_i \overset{\text{iid}}{\sim} \textsf{N}(0, \tau^2)\) for \(i = 1, \ldots, n\). Under this formulation, \(\sigma^2\) measures the variability in the global connectivity effect, while \(\tau^2\) captures the heterogeneity in individual tendencies to form connections among network actors. Additionally, the variance components are assigned inverse-gamma priors: \(\sigma^2 \sim \textsf{IG}(a_\sigma, b_\sigma)\) and \(\tau^2 \sim \textsf{IG}(a_\tau, b_\tau)\), where \(a_\sigma\), \(b_\sigma\), \(a_\tau\), and \(b_\tau\) are model hyperparameters (fixed values predefined by the analyst).

The choice of the Normal distribution is common in random effects models because it does not favor any specific direction in unobserved (latent) characteristics when such effects arise from multiple underlying factors, reflecting an assumption of no systematic bias. Moreover, despite the nonlinearity introduced by the link function, conjugacy can still be achieved through the introduction of auxiliary variables, simplifying the analysis of the posterior distribution. While the Normal distribution is a typical initial assumption, the model can be extended to accommodate more flexible distributions if the data suggest features such as asymmetry or heavy tails. Similarly, several alternative distributions to the inverse-gamma are available for modeling variance components, particularly when there is a need to prioritize specific ranges of variance values. Some of the most common options include the Uniform distribution and the Half-Cauchy distribution.

\begin{figure}[!htb]
\centering
\includegraphics[scale=0.8]{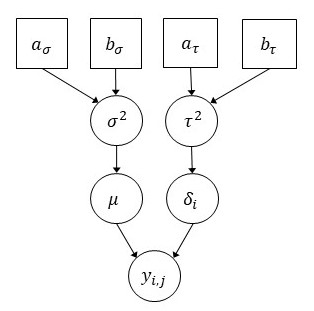}  
\caption{Directed Acyclic Graph (DAG) representing the sociality model.}
\label{fig_DAG}
\end{figure}

The set of parameters for this sociality model for undirected binary networks is defined as \(\Theta = \{\mu, \delta_1, \dots, \delta_n, \sigma^2, \tau^2\}\), comprising \(n + 3\) parameters. The number of parameters grows linearly with the number of nodes \(n\), ensuring computational efficiency, particularly for sparse networks with relatively few connections. This linear relationship prevents the model from becoming overly complex as the network size increases, making it well-suited for large but sparse networks. Additionally, the model includes a set of hyperparameters, \(\{a_\sigma, b_\sigma, a_\tau, b_\tau\}\), which define the priors for the variance components \(\sigma^2\) and \(\tau^2\), thus influencing the model's flexibility and regularization. The structure of this model is illustrated using a directed acyclic graph (DAG), shown in Figure \ref{fig_DAG}, which visually represents the dependencies and hierarchical structure of the model, clarifying the relationships between parameters and observed data.

\subsection{Model identifiability}

Although \(\mu\) and \(\delta_i\) are not completely interchangeable, the model may still face a problem of relative identifiability. Changes in \(\mu\) can be offset by adjustments in \(\delta_i\) in the opposite direction, preserving specific relationships between the interaction probabilities \(\theta_{i,j}\) and hindering unique parameter identification. For instance, let \(\mu' = \mu + k\) and \(\delta_i' = \delta_i - k/2\), for $i = 1,\ldots,n$. In this case, the linear predictor satisfies \(\mu' + \delta_i' + \delta_j' = \mu + \delta_i + \delta_j\), showing that distinct combinations of \(\mu\) and \(\delta_i\) can produce identical interaction probabilities \(\theta_{i,j}\).

Setting \(\delta_i \sim \textsf{N}(0, \tau^2)\) a priori does not fully resolve the model's identifiability issue. Although the \(\delta_i\) have a mean of zero under the prior, their sum is not guaranteed to be zero in the posterior. In practice, the \(\delta_i\) may collectively drift after incorporating the data, resulting in a compensatory effect with \(\mu\). To ensure that the constraint \(\sum_i \delta_i = 0\) is upheld during inference, it must be explicitly imposed within the model.

To ensure full identifiability, the most common approach is to impose the constraint \(\sum_i \delta_i = 0\). This establishes a shared reference point for the individual effects and ensures that \(\mu\) represents only the global average effect. In this work, we adopt this approach. Alternative methods include fixing \(\mu = 0\), which simplifies the model but may not be suitable if the global effect is of interest, or selecting a reference node (e.g., \(i = 1\)) and setting \(\delta_1 = 0\), allowing the \(\delta_i\) to be interpreted as relative effects with respect to the reference node. Each of these restrictions ensures model identifiability and provides a clear, unique interpretation of the parameters.

There are several methods to enforce the constraint \(\sum_i \delta_i = 0\). The most commonly used and straightforward approach involves directly applying the constraint during the sampling process in an MCMC algorithm or during parameter estimation in a variational algorithm through an explicit transformation at each iteration. This ensures that the \(\delta_i\) satisfy the restriction without requiring modifications to the priors or the model structure. This method is both practical and efficient, particularly in iterative implementations, and it is the approach adopted in this work. Alternative methods include redefining the \(\delta_i\) as deviations from their mean, \(\delta_i' = \delta_i - \bar{\delta}\), where \(\bar{\delta} = \frac{1}{n} \sum_i \delta_i\), or conditioning the prior distribution of the \(\delta_i\) on the sum being zero, $p(\boldsymbol{\delta} \mid \tau^2) \propto \prod_{i=1}^n \textsf{N}(\delta_i \mid 0, \tau^2) \cdot I\left(\textstyle\sum_i \delta_i = 0\right)$, which is equivalent to \(\boldsymbol{\delta} \sim \textsf{N}(\mathbf{0}_n, \tau^2 \mathbf{I}_n - \frac{\tau^2}{n} \mathbf{1}_n\mathbf{1}_n^\top)\).

\subsection{Prior elicitation}\label{sec_prior_elicitation}

The selection of hyperparameters for the priors of the global variance \(\sigma^2\) and the individual-specific variance \(\tau^2\) is critical for achieving optimal model performance, as the model's behavior is highly sensitive to these choices. These parameters regulate the uncertainty associated with the global connectivity effect \(\mu\) and the individual heterogeneity \(\delta_i\). A widely recommended guideline is to ensure a priori that \(\textsf{E}(\sigma^2) \leq \textsf{E}(\tau^2)\), reflecting the assumption that individual variability should dominate over the uncertainty in global connectivity.

A direct application of the double expectation theorem, the total variance theorem, and the properties of the Normal distribution reveals that the linear predictor \(\eta_{i,j} = \mu + \delta_i + \delta_j\) marginally follows a zero-mean Normal distribution with variance  
\[
\textsf{Var}(\eta_{i,j}) = \frac{b_\sigma}{a_\sigma - 1} + 2 \frac{b_\tau}{a_\tau - 1}.
\]  
This expression demonstrates how the global variance \(\sigma^2\) and the subject-specific variance \(\tau^2\) contribute to the overall variability of \(\eta_{i,j}\). Since \(\textsf{E}(\eta_{i,j}) = 0\), the model is a priori centered on an Erdős-Rényi network \citep{bollobas1998random} with an expected connection probability of approximately 0.5, leading to \(\textsf{E}(\theta_{i,j}) \approx 0.5\).

To select appropriate hyperparameters for the model, a sensible starting point is to set \(a_\sigma = a_\tau = 2\), resulting in a proper prior with a finite mean but an infinite coefficient of variation, reflecting a heavy-tailed distribution. Alternatively, setting \(a_\sigma = a_\tau = 3\) yields a coefficient of variation equal to 1, indicating more concentrated priors. Both choices allow for the possibility of large values of \(\sigma^2\) and \(\tau^2\) while maintaining flexibility in prior specification. 

When \(a_\sigma = a_\tau = 2\), the variance of the linear predictor is \(\textsf{Var}(\eta_{i,j}) = b_\sigma + 2b_\tau\). To ensure that \(\eta_{i,j}\) follows a standard Normal distribution a priori, we set \(\textsf{Var}(\eta_{i,j}) = 1\), which leads to approximately uniform interaction probabilities, as \(\Phi(Z) \sim \textsf{U}(0,1)\) when \(Z\) is standard Normal. Splitting \(b_\sigma + 2b_\tau = 1\) equally among all terms gives \(b_\sigma = b_\tau = 1/3\), resulting in \(\textsf{E}(\sigma^2) = \textsf{E}(\tau^2) = 1/3\). An alternative allocation is \(b_\sigma = 1/2\) and \(b_\tau = 1/4\), by dividing 1 equally between \(b_\sigma\) and \(2b_\tau\). Although this makes \(\textsf{E}(\sigma^2) > \textsf{E}(\tau^2)\), the interaction probabilities remain uniformly distributed, preserving the intended prior structure.

Alternative configurations for the hyperparameters include \((a, b) = (2, 1)\) or \((a, b) = (3, 2)\), both of which yield a unit expected value for the corresponding variance component. The first configuration results in an infinite coefficient of variation, reflecting a highly dispersed and heavy-tailed prior, while the second has a coefficient of variation equal to 1, representing a more concentrated prior. These choices are beneficial as they provide sufficient flexibility to account for variability without imposing overly restrictive assumptions. However, a limitation of this setup is that the interaction probabilities are not uniformly distributed a priori. Instead, they tend to favor extreme values, either low or high interaction probabilities, which could bias the model's a priori behavior.

\begin{figure}[!htb]
    \centering
    \subfigure[\(a_\sigma = a_\tau = 2\) and \(b_\sigma = b_\tau = 1/3\).]{
        \includegraphics[scale=0.42]{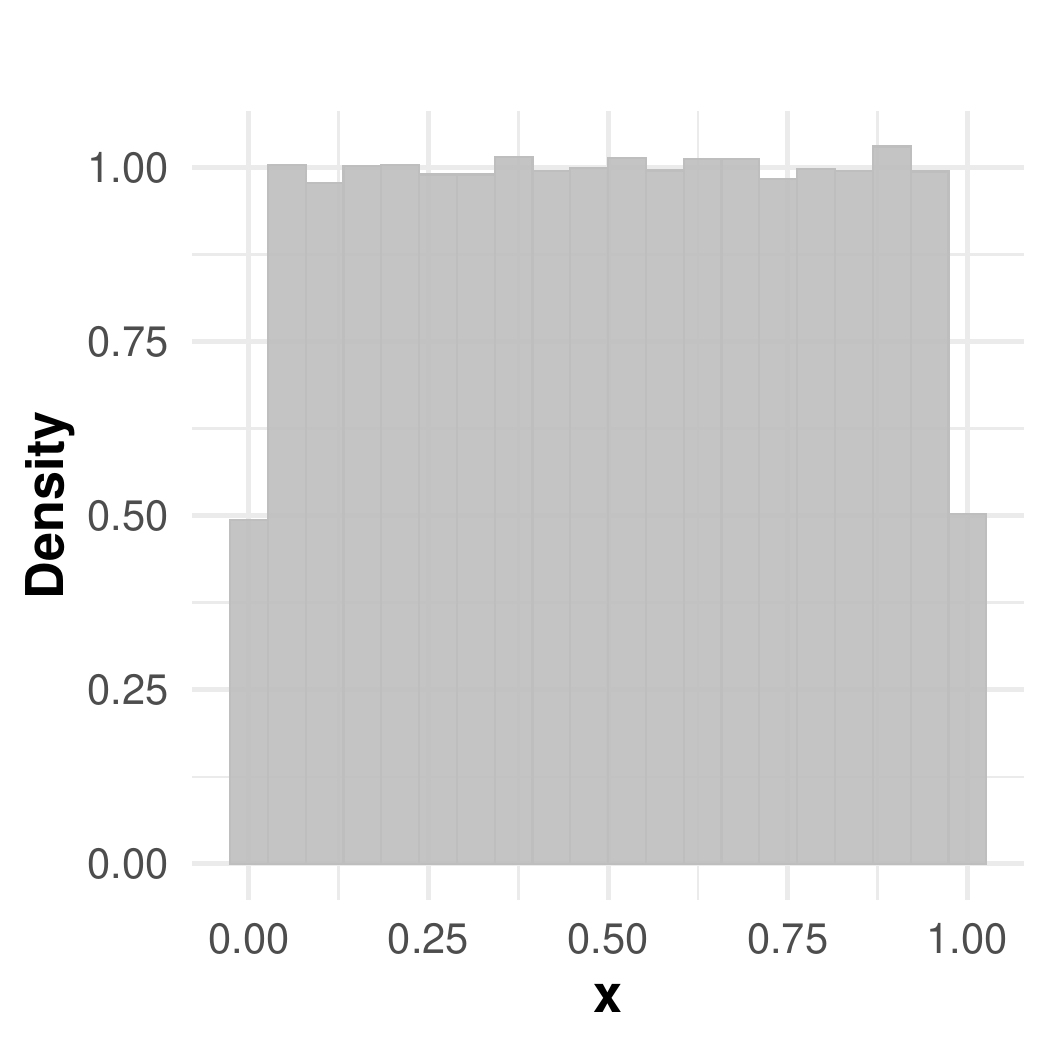}
    }
    \hspace{0em} 
    \subfigure[\(a_\sigma = a_\tau = 2\) and \(b_\sigma = b_\tau = 1\).]{
        \includegraphics[scale=0.42]{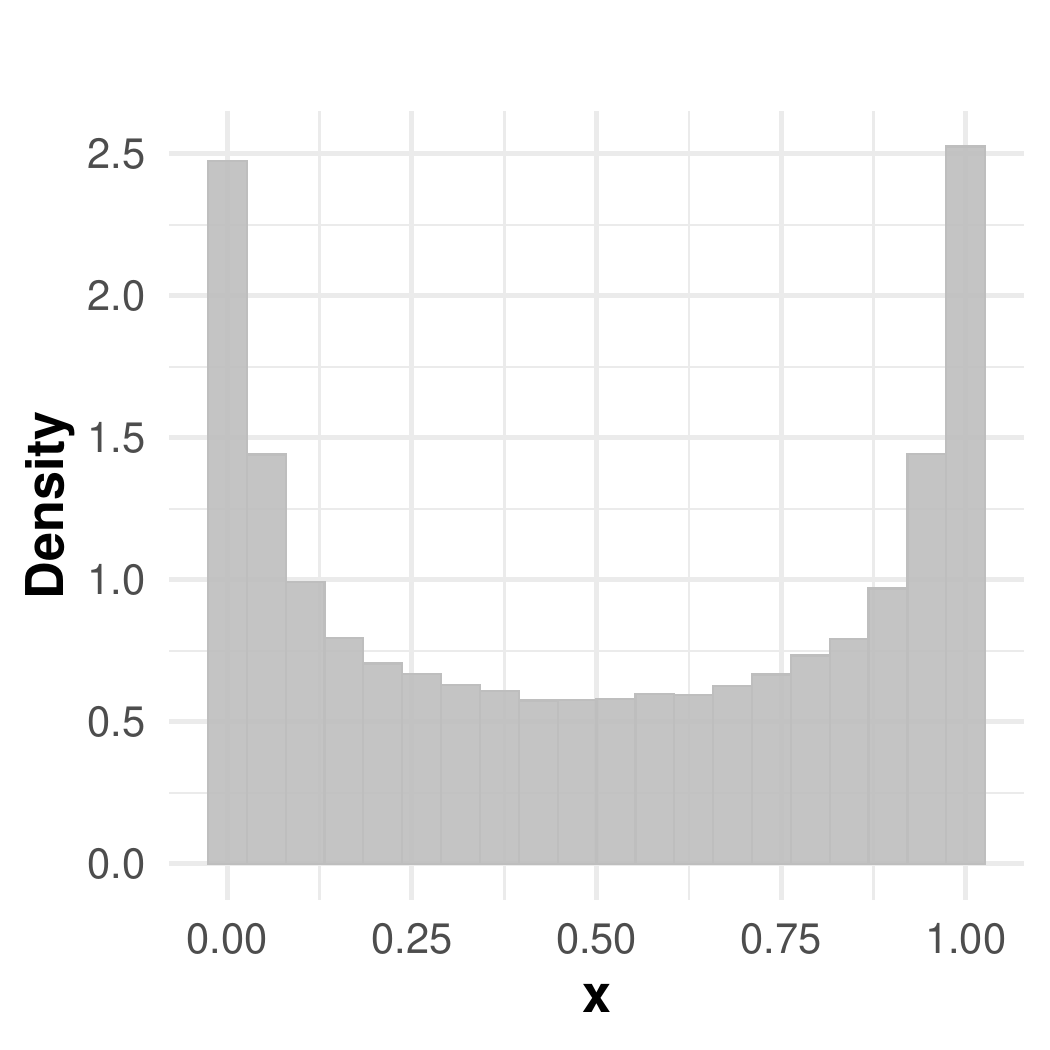}
    }
    \caption{Prior distribution of interaction probabilities under two different hyperparameter configurations.}
    \label{fig_prior_simulation}
\end{figure}

Figure \ref{fig_prior_simulation} presents histograms of 100,000 independent samples drawn from the induced marginal prior distribution of the interaction probabilities \(\theta_{i,j}\), under two different hyperparameter configurations. Panel (a) corresponds to \(a_\sigma = a_\tau = 2\) and \(b_\sigma = b_\tau = 1/3\), while panel (b) illustrates the case with \(a_\sigma = a_\tau = 2\) and \(b_\sigma = b_\tau = 1\). These histograms reflect how the choice of hyperparameters affects the a priori distribution of interaction probabilities, with panel (a) leading to a distribution closer to uniformity and panel (b) exhibiting a greater concentration around extreme probabilities, favoring values near 0 or 1.

\section{Computation}\label{sec_computation}

The posterior distribution of the model combines information from the observed data with prior beliefs about the model parameters, enabling the model to fit the evidence while adhering to the regularization imposed by the prior distributions. Derived directly from Bayes' theorem, the posterior distribution is expressed as \(p(\Theta \mid \mathbf{Y}) \propto p(\mathbf{Y} \mid \Theta) \, p(\Theta)\). Thus, the posterior distribution in this case is given by:  
\begin{align*}
p(\Theta \mid \mathbf{Y}) &\propto 
\prod_{i < j} \textsf{Ber}(y_{i,j} \mid \theta_{i,j}) \, \textsf{N}(\mu \mid 0, \sigma^2) \prod_{i=1}^n \textsf{N}(\delta_i \mid 0, \tau^2) \, \textsf{IG}(\sigma^2 \mid a_\sigma, b_\sigma) \, \textsf{IG}(\tau^2 \mid a_\tau, b_\tau) \\
&\propto
\prod_{i < j} \theta_{i,j}^{y_{i,j}} (1 - \theta_{i,j})^{1 - y_{i,j}} 
\times (\sigma^2)^{-1/2} \exp\left(-\frac{\mu^2}{2\sigma^2}\right)
\times \prod_{i=1}^n (\tau^2)^{-1/2} \exp\left(-\frac{\delta_i^2}{2\tau^2}\right)\\
&\quad\,\,
\times (\sigma^2)^{-(a_\sigma + 1)} \exp\left(-\frac{b_\sigma}{\sigma^2}\right)
\times (\tau^2)^{-(a_\tau + 1)} \exp\left(-\frac{b_\tau}{\tau^2}\right),
\end{align*}
where \(\theta_{i,j} = \Phi(\mu + \delta_i + \delta_j)\) represents the probability of a connection between actors \(i\) and \(j\), determined by the probit link function \(\Phi(\cdot)\).

\subsection{Markov chain Monte Carlo}

The posterior distribution \(p(\Theta \mid \mathbf{Y})\) can be explored using Markov chain Monte Carlo (MCMC; e.g., \citealt{gamerman2006markov}). This approach approximates the posterior by generating dependent but approximately identically distributed samples \(\Theta^{(1)}, \ldots, \Theta^{(S)}\), where \(\Theta^{(s)} = \{\mu^{(s)}, \delta_1^{(s)}, \dots, \delta_n^{(s)}, \sigma^{2\,(s)}, \tau^{2\,(s)}\}\) for \(s = 1, \ldots, S\). Point and interval estimates can then be obtained from the empirical distributions of these samples. The computational algorithm uses Gibbs sampling with Metropolis-Hastings steps, as the full conditional distributions for \(\mu\) and \(\delta_i\) lack closed forms due to the nonlinear relationship introduced by the probit link function between the parameters and the observations.

To facilitate computation, we follow \cite{albert1993bayesian} and introduce independent auxiliary random variables \(z_{i,j}\) such that  
\[
y_{i,j} = 
\begin{cases} 
1 & \text{if } z_{i,j} > 0, \\
0 & \text{if } z_{i,j} \leq 0,
\end{cases}
\]
where \(z_{i,j} \mid \mu, \delta_i, \delta_j \sim \textsf{N}(\mu + \delta_i + \delta_j, 1)\). If the model is formulated using a logit link function instead of a probit link, it is also possible to introduce auxiliary variables, but these would follow a Polya-Gamma distribution rather than a Normal distribution, as proposed by \cite{polson2013bayesian}.

The introduction of latent variables \(z_{i,j}\) preserves the original model because these variables are consistent with the underlying probability structure of the initial model. By integrating \(z_{i,j}\) over its conditional distribution, the original probability of \(y_{i,j}\) is recovered. Specifically, the probability of \(y_{i,j} = 1\) is equivalent to the probability that \(z_{i,j} > 0\):  \[
\textsf{P}(z_{i,j} > 0 \mid \mu, \delta_i, \delta_j) = \int_0^\infty \textsf{N}(z_{i,j} \mid \mu + \delta_i + \delta_j, 1) \, \textsf{d}z_{i,j} = \Phi(\mu + \delta_i + \delta_j) = \textsf{P}(y_{i,j} = 1 \mid \mu, \delta_i, \delta_j).
\]
This approach linearizes the model by introducing auxiliary variables \( z_{i,j} \) that follow a standard normal distribution, replacing the link function \( \Phi(\cdot) \). This reformulation simplifies the computation of full conditional distributions, ensuring they take standard forms, which facilitates the implementation of a Gibbs sampler.

The full conditional distribution (FCD) of the model parameters can be derived by examining the dependencies in the full posterior distribution. Consequently, we have that the FCDs are given by:
\begin{itemize}
    \item \( z_{i,j} \mid y_{i,j}, \mu, \delta_i, \delta_j \) follows a truncated Normal distribution:
    \[
    z_{i,j} \mid y_{i,j}, \mu, \delta_i, \delta_j \sim
    \begin{cases} 
    \textsf{TN}_{(0,\infty)}\,\,\, (\mu + \delta_i + \delta_j, 1), & \text{if } y_{i,j} = 1, \\
    \textsf{TN}_{(-\infty,0]}(\mu + \delta_i + \delta_j, 1), & \text{if } y_{i,j} = 0.
    \end{cases}
    \]

    \item \( \mu \mid \mathbf{z}, \boldsymbol{\delta}, \sigma^2 \sim \textsf{N}(M, V^2) \), with:
    \[
    M = V^2 \sum_{i<j} (z_{i,j} - \delta_i - \delta_j), \quad V^2 = \left(\frac{1}{\sigma^2} + \frac{n(n-1)}{2} \right)^{-1}.
    \]

    \item \( \delta_i \mid \mathbf{z}, \mu, \tau^2 \sim \textsf{N}(M_i, V_i^2) \), with:
    \[
    M_i = V_i^2 \sum_{j \neq i} (z_{i,j} - \mu - \delta_j), \quad V_i^2 = \left(\frac{1}{\tau^2} +  n-1 \right)^{-1}.
    \]

    \item \( \sigma^2 \mid \mu \sim \textsf{IG}(A, B) \), with:
    \[
    A = a_\sigma + \frac{1}{2}, \quad B = b_\sigma + \frac{\mu^2}{2}.
    \]

    \item \( \tau^2 \mid \boldsymbol{\delta} \sim \textsf{IG}(A, B) \), with:
    \[
    A = a_\tau + \frac{n}{2}, \quad B = b_\tau + \frac{\sum_{i} \delta_i^2}{2}.
    \]
\end{itemize}

Let \(\phi^{(s)}\) denote the state of the parameter \(\phi\) in the \(s\)-th iteration of the Gibbs sampling algorithm, for \(s = 1, \ldots, S\). The Gibbs sampling algorithm for obtaining samples from the posterior distribution \(p(\Theta \mid \mathbf{Y})\) proceeds as follows:
\begin{enumerate}
    \item \textbf{Initialization.} Initialize each parameter in the model with an initial configuration, 
    $$
    z^{(0)}_{1,2}, \ldots, z^{(0)}_{n-1,n}, \mu^{(0)}, \delta_1^{(0)}, \ldots, \delta_n^{(0)}, \sigma^{2\,(0)}, \tau^{2\,(0)}.
    $$

    \item \textbf{Iterative updates.} Iteratively update the parameters 
    $$
    z^{(s-1)}_{1,2}, \ldots, z^{(s-1)}_{n-1,n}, \mu^{(s-1)}, \delta_1^{(s-1)}, \ldots, \delta_n^{(s-1)}, \sigma^{2\,(s-1)}, \tau^{2\,(s-1)},
    $$ 
    as follows:
    \begin{enumerate}
        \item[2.1] Sample \(z^{(s)}_{i,j}\) from $p(z_{i,j} \mid y_{i,j}, \mu^{(s-1)}, \delta_i^{(s-1)}, \delta_j^{(s-1)})$.
        \item[2.2] Sample \(\mu^{(s)}\) from  $p(\mu \mid \mathbf{z}^{(s)}, \boldsymbol{\delta}^{(s-1)}, \sigma^{2\,(s-1)})$.
        \item[2.3] Sample \(\delta_i^{(s)}\) from $p(\delta_i \mid \mathbf{z}^{(s)}, \mu^{(s)}, \tau^{2\,(s-1)})$.
        \item[2.4] Compute $\bar\delta = \frac{1}{n}\sum_i \delta_i$. \\ Update $\delta_i \leftarrow \delta_i - \bar{\delta}$.
        \item[2.5] Sample \(\sigma^{2\,(s)}\) from $p(\sigma^2 \mid \mu^{(s)})$.
        \item[2.6] Sample \(\tau^{2\,(s)}\) from $p(\tau^2 \mid \boldsymbol{\delta}^{(s)})$.
    \end{enumerate}

    \item \textbf{Cycle.} Repeat the above steps until convergence is achieved.
\end{enumerate}

\subsection{Variational inference}

Variational Inference (VI; e.g., \citealt{beal2003variational}, \citealt{ormerod2010explaining}, \citealt{blei2017variational}) is a deterministic optimization method for approximating posterior distributions in Bayesian models. Unlike MCMC, which relies on sampling from the posterior \(p(\Theta \mid \mathbf{Y})\), VI approximates \(p(\Theta \mid \mathbf{Y})\) with a simpler probability distribution \(q(\Theta; \boldsymbol{\lambda})\), where \(\lambda\) denotes the variational parameters. The objective is to find the optimal \(\lambda\) such that \(q(\Theta; \boldsymbol{\lambda})\) closely approximates the true posterior \(p(\Theta \mid \mathbf{Y})\).

It can be shown (see Appendix \ref{app_decomposition}) that the marginal log-likelihood \(\log p(\mathbf{Y})\) can be expressed as:  
\[
\log p(\mathbf{Y}) = \int q(\Theta; \boldsymbol{\lambda}) \log \frac{p(\mathbf{Y}, \Theta)}{q(\Theta; \boldsymbol{\lambda})} \, \textsf{d}\Theta + \int q(\Theta; \boldsymbol{\lambda}) \log \frac{q(\Theta; \boldsymbol{\lambda})}{p(\Theta \mid \mathbf{Y})} \, \textsf{d}\Theta.
\]
The first term on the right-hand side is known as the Evidence Lower Bound (ELBO):
\[
\text{ELBO} = \textsf{E}_{q(\Theta; \boldsymbol{\lambda})}(\log p(\mathbf{Y}, \Theta)) - \textsf{E}_{q(\Theta; \boldsymbol{\lambda})}(\log q(\Theta; \boldsymbol{\lambda})),
\]
while the second term is the Kullback-Leibler (KL) divergence, which quantifies the discrepancy between \(q(\Theta; \boldsymbol{\lambda})\) and \(p(\Theta \mid \mathbf{Y})\):
\[
\text{KL}(q(\Theta; \boldsymbol{\lambda}) \| p(\Theta \mid \mathbf{Y})) = \int q(\Theta; \boldsymbol{\lambda}) \log \frac{q(\Theta; \boldsymbol{\lambda})}{p(\Theta \mid \mathbf{Y})} \, \textsf{d}\Theta.
\]
VI minimizes the KL divergence indirectly by maximizing the ELBO, which serves as a lower bound on the marginal log-likelihood \(\log p(\mathbf{Y})\). 

Defining the variational family involves selecting a distribution \(q(\Theta; \boldsymbol{\lambda})\) that achieves a balance between flexibility and computational efficiency. The chosen family must be flexible enough to closely approximate the true posterior \(p(\Theta \mid \mathbf{Y})\) while allowing for tractable optimization procedures. A widely used approach is the mean-field assumption:  
\[
q(\Theta; \boldsymbol{\lambda}) = \prod_{i} q(\theta_i; \lambda_i),
\]  
which simplifies the variational distribution by factorizing it into independent components. Despite its simplicity, the mean-field assumption often provides a sufficiently accurate approximation for many practical applications, while retaining the ability to capture key features of complex posterior distributions.

VI optimization iteratively refines the variational parameters \(\boldsymbol{\lambda}\) to maximize the ELBO. A widely used approach for this optimization is the Coordinate Ascent Variational Inference (CAVI) method. This deterministic technique maximizes the ELBO by sequentially updating each component of the variational distribution while keeping the others fixed. For each component \(q(\theta_i; \lambda_i)\), the update is performed by solving:
\[
\log q(\theta_i; \lambda_i) \propto \textsf{E}_{q(\Theta \setminus \theta_i)}(\log p(\mathbf{Y}, \Theta)),
\]
where \(\Theta \setminus \theta_i\) denotes all parameters except \(\theta_i\). Convergence is ensured by tracking the ELBO across iterations, allowing the algorithm to iteratively refine the approximation until it stabilizes at a local maximum of the ELBO.

Assuming a mean-field variational family of the form
\[
q(\Theta) = \prod_{i < j} q(z_{i,j}) \cdot q(\mu) \cdot \prod_{i=1}^n q(\delta_i) \cdot q(\sigma^2) \cdot q(\tau^2),
\]
the CAVI updates for each variational factor are derived sequentially by optimizing the ELBO with respect to each factor while holding the others fixed. These updates are given by:
\begin{itemize}
    \item  \( q(z_{i,j}) \) is a truncated Normal distribution:
    \[
    q(z_{i,j}) = 
    \begin{cases} 
    \textsf{TN}_{(0,\infty)}\,\,\, (\textsf{E}_{q(\mu)}(\mu) + \textsf{E}_{q(\delta_i)}(\delta_i) + \textsf{E}_{q(\delta_j)}(\delta_j), 1), & \text{if } y_{i,j} = 1, \\
    \textsf{TN}_{(-\infty,0]}(\textsf{E}_{q(\mu)}(\mu) + \textsf{E}_{q(\delta_i)}(\delta_i) + \textsf{E}_{q(\delta_j)}(\delta_j), 1), & \text{if } y_{i,j} = 0.
    \end{cases}
    \]

    \item $q(\mu) = \textsf{N}(\mu_\mu,\sigma^2_\mu)$, with:
    \[
    \sigma_\mu^2 = \left(\textsf{E}_{q(\sigma^2)}\left(\frac{1}{\sigma^2}\right) + \frac{n(n-1)}{2}\right)^{-1}, \,\,\,
    \mu_\mu = \sigma_\mu^2 \sum_{i<j} \left( \textsf{E}_{q(z_{i,j})}(z_{i,j}) - \textsf{E}_{q(\delta_i)}(\delta_i) - \textsf{E}_{q(\delta_j)}(\delta_j) \right).
    \]

    \item $q(\delta_i) = \textsf{N}(\mu_{\delta_i}, \sigma_{\delta_i}^2)$, with:
    \[
    \sigma_{\delta_i}^2 = \left(\textsf{E}_{q(\tau^2)}\left(\frac{1}{\tau^2}\right) + n-1 \right)^{-1},\,\,\,
    \mu_{\delta_i} = \sigma_{\delta_i}^2 \sum_{j \neq i} \left( \textsf{E}_{q(z_{i,j})}(z_{i,j}) - \textsf{E}_{q(\mu)}(\mu) - \textsf{E}_{q(\delta_j)}(\delta_j) \right).
    \]

    \item $q(\sigma^2) = \textsf{IG}(\alpha_\sigma, \beta_\sigma)$, with:
    \[
    \alpha_\sigma = a_\sigma + \frac{1}{2}, \quad \beta_\sigma = b_\sigma + \frac{\textsf{E}_{q(\mu)}(\mu^2)}{2}.
    \]

    \item $q(\tau^2) = \textsf{IG}(\alpha_\tau, \beta_\tau)$, with:
    \[
    \alpha_\tau = a_\tau + \frac{n}{2}, \quad \beta_\tau = b_\tau + \frac{\sum_i \textsf{E}_{q(\delta_i)}(\delta_i^2)}{2}.
    \]
\end{itemize}

The algorithm for the CAVI applied to the sociality model, detailing the updates for each variational parameter, can be outlined as follows:
\begin{enumerate}
    \item \textbf{Initialization.} Initialize the variational parameters 
    $$
    \mu_{z_{1,2}},\ldots,\mu_{z_{n-1,n}}, \mu_\mu, \sigma_\mu^2, \mu_{\delta_1}, \sigma_{\delta_1}^2,\ldots,\mu_{\delta_n}, \sigma_{\delta_n}^2, \alpha_\sigma, \beta_\sigma, \alpha_\tau, \beta_\tau
    $$
    and set a convergence tolerance \(\epsilon\) (e.g., \(10^{-6}\)).

    \item \textbf{Iterative updates.} Repeat the following updates until convergence:
    \begin{enumerate}
        \item[2.1] Compute
        $$
        m_{z_{i,j}} = \mu_\mu + \mu_{\delta_i} + \mu_{\delta_j}.
        $$

        Update
        $$
        \mu_{z_{i,j}} \leftarrow
        m_{z_{i,j}} + \frac{\phi(-m_{z_{i,j}})}{\Phi(-m_{z_{i,j}})},
        $$
        if $y_{i,j} = 1$, and 
        $$
        \mu_{z_{i,j}} \leftarrow
        m_{z_{i,j}} - \frac{\phi(-m_{z_{i,j}})}{1-\Phi(-m_{z_{i,j}})},
        $$
        if $y_{i,j} = 0$.
        
        Bound
        $$
        \mu_{z_{i,j}} \leftarrow \min(c, \max(-c, \mu_{z_{i,j}})),
        $$
        for fixed $c$.
        \item[2.2] Update
        $$
        \sigma_\mu^2 \leftarrow \left(\frac{\alpha_\sigma}{\beta_\sigma} + \frac{n(n-1)}{2} \right)^{-1}, \quad 
        \mu_\mu \leftarrow \sigma_\mu^2 \sum_{i<j} \left(\mu_{z_{i,j}} - \mu_{\delta_i} - \mu_{\delta_j}\right).
        $$
        \item[2.3] Update
        $$
        \sigma_{\delta_i}^2 \leftarrow \left(\frac{\alpha_\tau}{\beta_\tau} + n-1 \right)^{-1}, \quad 
        \mu_{\delta_i} \leftarrow \sigma_{\delta_i}^2 \sum_{j \neq i} \left(\mu_{z_{i,j}} - \mu_\mu - \mu_{\delta_j}\right).
        $$
        \item[2.5] Compute 
        $$
        \bar\mu_\delta = \frac{1}{n}\sum_i \mu_{\delta_i}.
        $$ 
        
        Update 
        $$
        \mu_{\delta_i} \leftarrow \mu_{\delta_i} - \bar\mu_\delta.
        $$
        
        \item[2.5] Update
        $$
        \alpha_\sigma \leftarrow a_\sigma + \frac12, \quad \beta_\sigma \leftarrow b_\sigma + \frac{\mu_\mu^2 + \sigma_\mu^2}{2}.
        $$
        \item[2.6] Update: 
        $$
        \alpha_\tau \leftarrow a_\tau + \frac{n}{2}, \quad \beta_\tau \leftarrow b_\tau + \frac{\sum_i (\mu_{\delta_i}^2 + \sigma_{\delta_i}^2)}{2}.
        $$
    \end{enumerate}
    \item \textbf{Convergence check.} 
    \begin{enumerate}
        \item[3.1] Compute $\text{ELBO} = \textsf{P} - \textsf{Q}$, where 
        $$\textsf{P} = \textsf{E}_{q(\Theta;\boldsymbol{\lambda})}(\log p(\mathbf{Y}, \Theta))
        \quad\text{and}\quad
        \textsf{Q} = \textsf{E}_{q(\Theta;\boldsymbol{\lambda})}(\log q(\Theta)).
        $$ 
        These components are computed as follows:
        \begin{align*}
        \textsf{P} &= -\tfrac{1}{2} \textstyle\sum_{i<j} ( \sigma_\mu^2 + \sigma_{\delta_i}^2 + \sigma_{\delta_j}^2 + (\mu_{z_{i,j}} - \mu_\mu - \mu_{\delta_i} - \mu_{\delta_j})^2 + \log 2\pi ) \\
        &\quad -\tfrac{1}{2} ( ( \mu_\mu^2 + \sigma_\mu^2) (\alpha_\sigma/\beta_\sigma) + \psi(\alpha_\sigma) - \log \beta_\sigma + \log 2\pi ) \\
        &\quad -\tfrac{1}{2} \textstyle\sum_{i} ( (\mu_{\delta_i}^2 + \sigma_{\delta_i}^2) (\alpha_\tau/\beta_\tau) + \psi(\alpha_\tau) - \log \beta_\tau + \log 2\pi ) \\
        &\quad +a_\sigma \log b_\sigma - \log \Gamma(a_\sigma) - (a_\sigma + 1)(\psi(\alpha_\sigma) - \log \beta_\sigma) - b_\sigma \, \alpha_\sigma/\beta_\sigma \\
        &\quad +a_\tau \log b_\tau - \log \Gamma(a_\tau) - (a_\tau + 1)(\psi(\alpha_\tau) - \log \beta_\tau) - b_\tau \, \alpha_\tau/\beta_\tau,
        \end{align*}
        and
        \begin{align*}
        \textsf{Q} &= \textsf{q}_{i,j} - \tfrac{1}{2}\log2\pi \\
        &\quad -\tfrac{1}{2} ( \log(2\pi \sigma_\mu^2) + 1  ) -\tfrac{n}{2}(\log(2\pi) + 1) - \tfrac{1}{2} \textstyle\sum_{i} \log \sigma_{\delta_i}^2 \\
        &\quad +\alpha_\sigma \log \beta_\sigma - \log \Gamma(\alpha_\sigma) - (\alpha_\sigma + 1)(\psi(\alpha_\sigma) - \log \beta_\sigma) - \alpha_\sigma \\
        &\quad +\alpha_\tau \log \beta_\tau - \log \Gamma(\alpha_\tau) - (\alpha_\tau + 1)(\psi(\alpha_\tau) - \log \beta_\tau) - \alpha_\tau,
        \end{align*}
        with
        $$
        \textsf{q}_{i,j}= 
        -\frac{1}{2} \left(1 + \frac{-\mu_{z_{i,j}} \phi(-\mu_{z_{i,j}})}{1 - \Phi(-\mu_{z_{i,j}})} - \left(\frac{\phi(-\mu_{z_{i,j}})}{1 - \Phi(-\mu_{z_{i,j}})}\right)^2\right) - \log(1 - \Phi(-\mu_{z_{i,j}})),
        $$
        if $y_{i,j} = 1$, and 
        $$
        \textsf{q}_{i,j}=
        -\frac{1}{2} \left(1 + \frac{-\mu_{z_{i,j}} \phi(-\mu_{z_{i,j}})}{\Phi(-\mu_{z_{i,j}})} - \left(\frac{-\phi(-\mu_{z_{i,j}})}{\Phi(-\mu_{z_{i,j}})}\right)^2\right) - \log(\Phi(-\mu_{z_{i,j}})),
        $$
        if $y_{i,j} = 0$.
        \item[3.2] Compute the change in ELBO: $\Delta_{\text{ELBO}} = \text{ELBO}_{\text{new}} - \text{ELBO}_{\text{old}}$.
        \item[3.3] If \(\Delta_{\text{ELBO}} < \epsilon\), stop the algorithm.
    \end{enumerate} 
\end{enumerate}
Here, \(\Gamma(\cdot)\) is the gamma function, \(\psi(\cdot)\) is the digamma function, \(\phi(\cdot)\) is the probability density function (PDF) of the standard Normal distribution, and \(\Phi(\cdot)\) is the cumulative distribution function (CDF) of the standard Normal distribution. The derivation of the ELBO is provided in detail in Appendix \ref{app_elbo}.

Bounding \(\textsf{E}_{q(z_{i,j})}(z_{i,j})\) is a practical solution to stabilize variational updates and prevent extreme oscillations. The mean of the truncated Normal, \(\textsf{E}_{q(z_{i,j})}(z_{i,j})\), involves ratios like \(\phi(-m_{z_{i,j}})/\Phi(-m_{z_{i,j}})\), which become numerically unstable for large \(|m_{z_{i,j}}|\), leading to unreliable updates. A fixed range \((-c, c)\), where \(c > 0\), ensures \(\textsf{E}_{q(z_{i,j})}(z_{i,j})\) stays within a global threshold, regardless of the scale of \(m_{z_{i,j}}\). The value of \(c\) can be set based on practical needs or model properties. This approach is simple, efficient, and does not require dynamic adjustments, making it robust for extreme cases. The update modifies \(\textsf{E}_{q(z_{i,j})}(z_{i,j})\) by clipping it to \((-c, c)\), ensuring numerical stability and maintaining algorithm reliability. For applications involving logistic or probit links, $c=3$ is often sufficient, as values outside this range have negligible contributions to the probabilities.

Finally, the output of the CAVI algorithm comprises the variational parameters that define the approximate posterior distributions for the model parameters \(\mathbf{z}\), \(\mu\), \(\boldsymbol{\delta}\), \(\sigma^2\), and \(\tau^2\). Specifically, it includes the means of the truncated Normal distributions for the latent variables \(z_{i,j}\), the means and variances for the Normal approximations of \(\mu\) and \(\boldsymbol{\delta}\), and the shape and rate parameters for the inverse-gamma approximations of \(\sigma^2\) and \(\tau^2\). These outputs allow for the computation of posterior means, variances, and credible intervals for each parameter, facilitating comprehensive probabilistic inference. The iterative evaluation of the ELBO ensures the algorithm's convergence while providing a quantitative measure of the quality of the approximation.

\section{Latent space models}\label{sec_latent_space_models}

This section outlines key aspects of latent space models, which provide a baseline for comparing the sociality model in the illustrations section. Random effects in generalized linear models effectively capture network structures. For conditionally independent $y_{i,j}$, the interaction probabilities are:  
\[
\textsf{Pr}(y_{i,j}=1\mid \boldsymbol{\beta}, \gamma_{i,j}, \boldsymbol{x}_{i,j} ) = g^{-1}(\boldsymbol{x}^\top_{i,j}\boldsymbol{\beta} + \gamma_{i,j}),\qquad i<j,
\]  
where $\boldsymbol{\beta}=(\beta_1,\ldots,\beta_P)$ represents fixed effects, $\boldsymbol{x}^\top_{i,j}\boldsymbol{\beta}$ captures patterns from covariates $\boldsymbol{x}_{i,j}$, $\gamma_{i,j}$ accounts for unobserved effects, and $g(\cdot)$ is a known link function. As noted in \citet{hoff-2008}, a jointly exchangeable random effects matrix $[\gamma_{i,j}]$ can be expressed as $\gamma_{i,j} = \alpha(\boldsymbol{u}_i, \boldsymbol{u}_j)$, where $\boldsymbol{u}_1, \ldots, \boldsymbol{u}_n$ are independent latent variables. The function $\alpha(\cdot,\cdot)$ is central to modeling relational data. Several formulations for $\alpha(\cdot,\cdot)$ have been proposed in the literature. The most relevant approaches are discussed below. A comprehensive review of these and various other models is presented in \citet{sosa2021review}.

\subsection{Distance models}

\citet{hoff-2002} propose that each actor \(i\) occupies a position \(\boldsymbol{u}_i \in \mathbb{R}^K\) in a Euclidean space, where the probability of an edge increases as actors become closer. This is modeled as \(\alpha(\boldsymbol{u}_i, \boldsymbol{u}_j) = -\|\boldsymbol{u}_i - \boldsymbol{u}_j\|\), where \(\|\cdot\|\) denotes the Euclidean norm. Such distance-based structures naturally induce homophily (stronger ties between nodes with similar characteristics) and enable effective visualization of social networks. However, these models may be less suitable for networks with high clustering.

We consider the following model:  
\[
y_{i,j} \mid \zeta, \boldsymbol{u}_i, \boldsymbol{u}_j \overset{\text{ind}}{\sim} \textsf{Ber}\left(\Phi\left(\zeta - \|\boldsymbol{u}_i - \boldsymbol{u}_j\|\right)\right),
\]  
where $\zeta$ represents the average edge propensity, and $\boldsymbol{u}_1, \ldots, \boldsymbol{u}_I$ are unobserved positions in $\mathbb{R}^K$. For Bayesian inference, we assign priors: $\zeta \mid \omega^2 \sim \textsf{N}(0, \omega^2)$ and $\boldsymbol{u}_i \mid \sigma^2 \overset{\text{iid}}{\sim} \textsf{N}(\boldsymbol{0}, \sigma^2 \mathbf{I})$, where $\mathbf{I}$ is the identity matrix. Hyperpriors are specified as $\omega^2 \sim \textsf{IG}(a_\omega, b_\omega)$ and $\sigma^2 \sim \textsf{IG}(a_\sigma, b_\sigma)$. To achieve reliable model performance, we set \(a_\omega = a_\sigma = a_\kappa = 3\) and \(b_\omega = b_\sigma = b_\kappa = 2\), resulting in weakly informative priors that have been empirically effective.

\subsection{Class models}

\citet{nowicki-2001} propose that each actor \(i\) belongs to a latent class \(u_i \in \{1, \ldots, K\}\), with relationships between classes defined by \(\alpha(u_i, u_j) = \theta_{\phi(u_i, u_j)}\). Here, \(\Theta = [\theta_{k,\ell}]\) is a symmetric \(K \times K\) matrix with \(0 < \theta_{k,\ell} < 1\), and \(\phi(u,v) = (\min\{u,v\}, \max\{u,v\})\). Known as stochastic block models (SBMs), these effectively capture stochastic equivalence, grouping nodes with similar relationship patterns. However, SBMs may struggle with actors that do not fit neatly into clusters \cite{hoff-2002}.

Class models group similar actors into clusters and model the probability of an edge between two actors based on their classes:
\[
y_{i,j} \mid u_i, u_j, \{\eta_{k,\ell}\} \overset{\text{ind}}{\sim} \textsf{Ber}\left( \Phi(\eta_{\phi(u_i, u_j)}) \right),
\]
where \(\boldsymbol{u} = (u_1, \ldots, u_n)\) are unobserved cluster indicators in \(\{1, \ldots, K\}\), and \(K\) is the fixed number of classes. The function \(\phi(a,b) = (\min\{a,b\}, \max\{a,b\})\) ensures symmetry in the community parameters, as \(\mathbf{Y}\) is symmetric. Actors \(i\) and \(j\) belong to the same class if \(u_i = u_j\). The community parameters \(\boldsymbol{\eta} = \{ \eta_{k,\ell} : k, \ell = 1, \ldots, K, k \leq \ell \}\) follow the prior \(\eta_{k,\ell} \mid \zeta, \tau^2 \overset{\text{iid}}{\sim} \textsf{N}(\zeta, \tau^2)\). 

Following standard practice in community detection, the entries of $\boldsymbol{u}$ are typically assumed to follow a categorical distribution on $\{1, \ldots, K\}$, with $\textsf{P}(u_i = k \mid \omega_k) = \omega_k$, for $k = 1, \ldots, K$. The probability vector $\boldsymbol{\omega} = (\omega_1, \ldots, \omega_K)$ satisfies $\sum_{k=1}^K \omega_k = 1$ and follows the prior $\boldsymbol{\omega} \mid \alpha \sim \textsf{Dir}\left(\tfrac{\alpha}{K}, \ldots, \tfrac{\alpha}{K}\right)$. The model is completed by assigning independent priors: $\zeta \sim \text{N}(\mu_\zeta, \sigma^2_\zeta)$, $\tau^2 \sim \textsf{IG}(a_\tau, b_\tau)$, and $\alpha \sim \textsf{G}(a_\alpha, b_\alpha)$, where $\mu_\zeta$, $\sigma^2_\zeta$, $a_\tau$, $b_\tau$, $a_\alpha$, and $b_\alpha$ are hyperparameters. To ensure appropriate model performance, we set $\mu_\zeta = 0$, $\sigma_\zeta^2 = 3$, $a_\tau = 3$, and $b_\tau = 2$, centering the prior interaction probabilities $\Phi(\eta_{k,\ell})$ around 0.5 with reasonable variability on the logit scale. Additionally, $a_\alpha = 1$ and $b_\alpha = 1$ define a diffuse prior for $\alpha$ centered at 1.

\subsection{Eigen models}

\citet{hoff-2008} and \citet{hoff-2009} model relationships between nodes as the weighted inner product of latent vectors \(\boldsymbol{u}_i \in \mathbb{R}^K\), with latent effects \(\alpha(\boldsymbol{u}_i, \boldsymbol{u}_j) = \boldsymbol{u}_i^\top \mathbf{\Lambda} \boldsymbol{u}_j\), where \(\mathbf{\Lambda}\) is a \(K \times K\) diagonal matrix. Referred to as the eigen model, this approach builds upon latent class and latent distance models by efficiently representing network features and accommodating both positive and negative homophily. Additionally, stochastically equivalent nodes may or may not form strong relationships \citep{hoff-2008}.

In the eigen model, the sampling distribution is given by
\[
y_{i,j} \mid \zeta, \boldsymbol{u}_{i}, \boldsymbol{u}_{j}, \mathbf{\Lambda} \overset{\text{ind}}{\sim} \textsf{Ber}\left( \Phi\left(\zeta + \boldsymbol{u}_{i}^\top \mathbf{\Lambda} \boldsymbol{u}_{j} \right) \right)\,,
\]  
where \(\boldsymbol{u}_i = (u_{i,1}, \ldots, u_{i,K})\) is a vector of latent characteristics in \(\mathbb{R}^K\) and \(\mathbf{\Lambda} = \text{diag}(\lambda_1, \ldots, \lambda_K)\) is a diagonal matrix of size \(K \times K\). This implies that \(\boldsymbol{u}_{i}^\top \mathbf{\Lambda} \boldsymbol{u}_{j} = \sum_{k=1}^K \lambda_k u_{i,k} u_{j,k}\), a quadratic form where \(\lambda_k\) weights the contribution of each latent dimension (positively or negatively) to the plausibility of observing an edge between actors \(i\) and \(j\). Following the same prior formulation used for distance models, we let \(\zeta \mid \omega^2 \sim \textsf{N}(0, \omega^2)\) and \(\boldsymbol{u}_i \mid \sigma^2 \overset{\text{iid}}{\sim} \textsf{N}(\boldsymbol{0}, \sigma^2 \mathbf{I})\), along with \(\omega^2 \sim \textsf{IG}(a_\omega, b_\omega)\) and \(\sigma^2 \sim \textsf{IG}(a_\sigma, b_\sigma)\). We complete the specification by assuming that \(\lambda_k \mid \kappa^2 \overset{\text{iid}}{\sim} \textsf{N}(0, \kappa^2)\), where \(\kappa^2 \sim \textsf{IG}(a_{\kappa}, b_{\kappa})\). Lastly, vaguely uninformative priors that have proven effective in practice are obtained by setting \(a_\omega = a_\sigma = a_\kappa = 3\) and \(b_\omega = b_\sigma = b_\kappa = 2\).

\section{Illustrations}

Section 5 evaluates the sociality model through empirical analyses and simulations. It first examines real-world networks, particularly the Lazega law firm dataset, assessing goodness-of-fit and predictive accuracy while demonstrating the model’s ability to capture degree heterogeneity and structural patterns. A simulation study then compares MCMC and VI in terms of goodness-of-fit, accuracy, and scalability, highlighting their trade-offs in estimating latent structures across different network settings. Finally, all the \texttt{R} and \texttt{C$^{++}$} code to replicate the results of this section is available at \url{https://github.com/jstats1702}.

\subsection{Collaborative working relationships}

\begin{figure}[!htb]
    \centering
    \subfigure[Adjacency matrix.]{
        \includegraphics[scale=0.42]{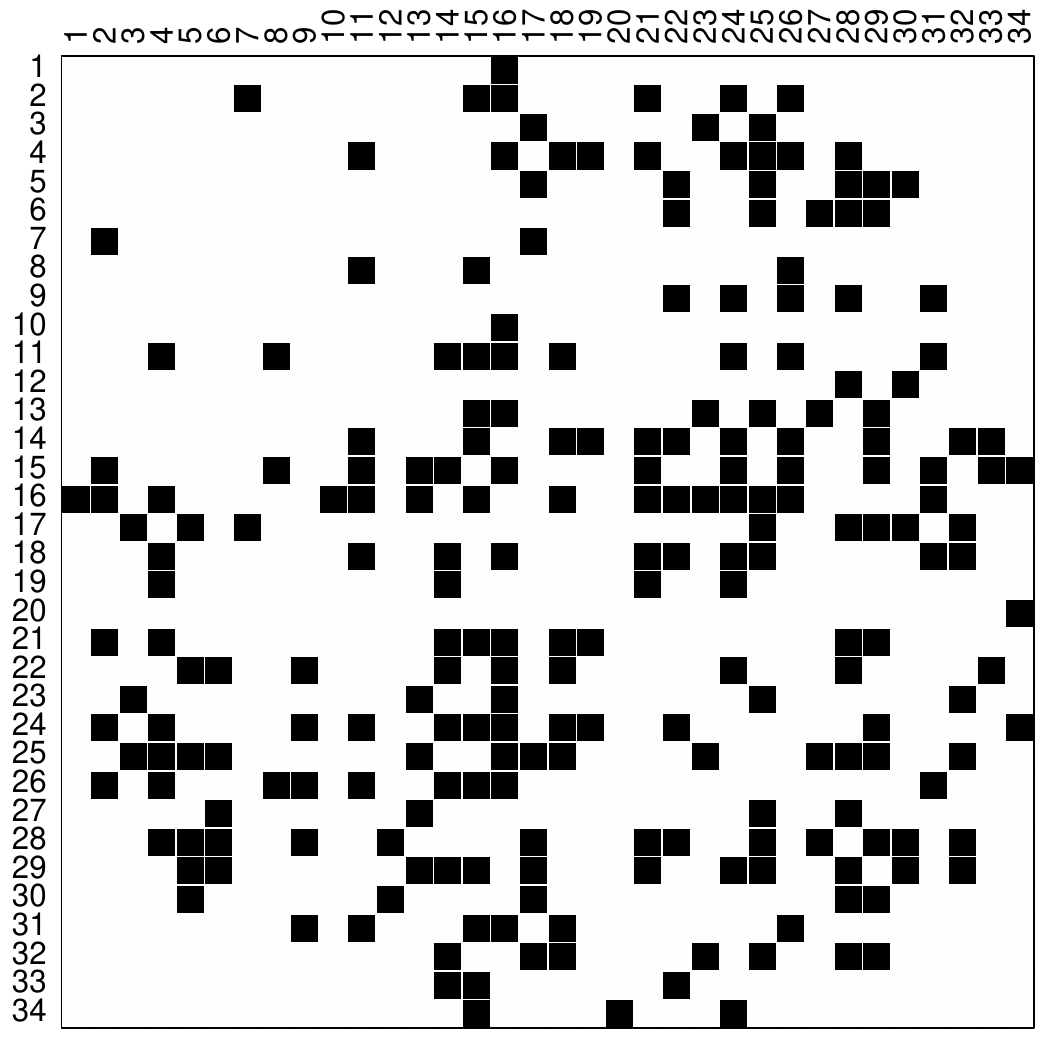}
    }
    \hspace{0em} 
    \subfigure[Graph.]{
        \includegraphics[scale=0.42]{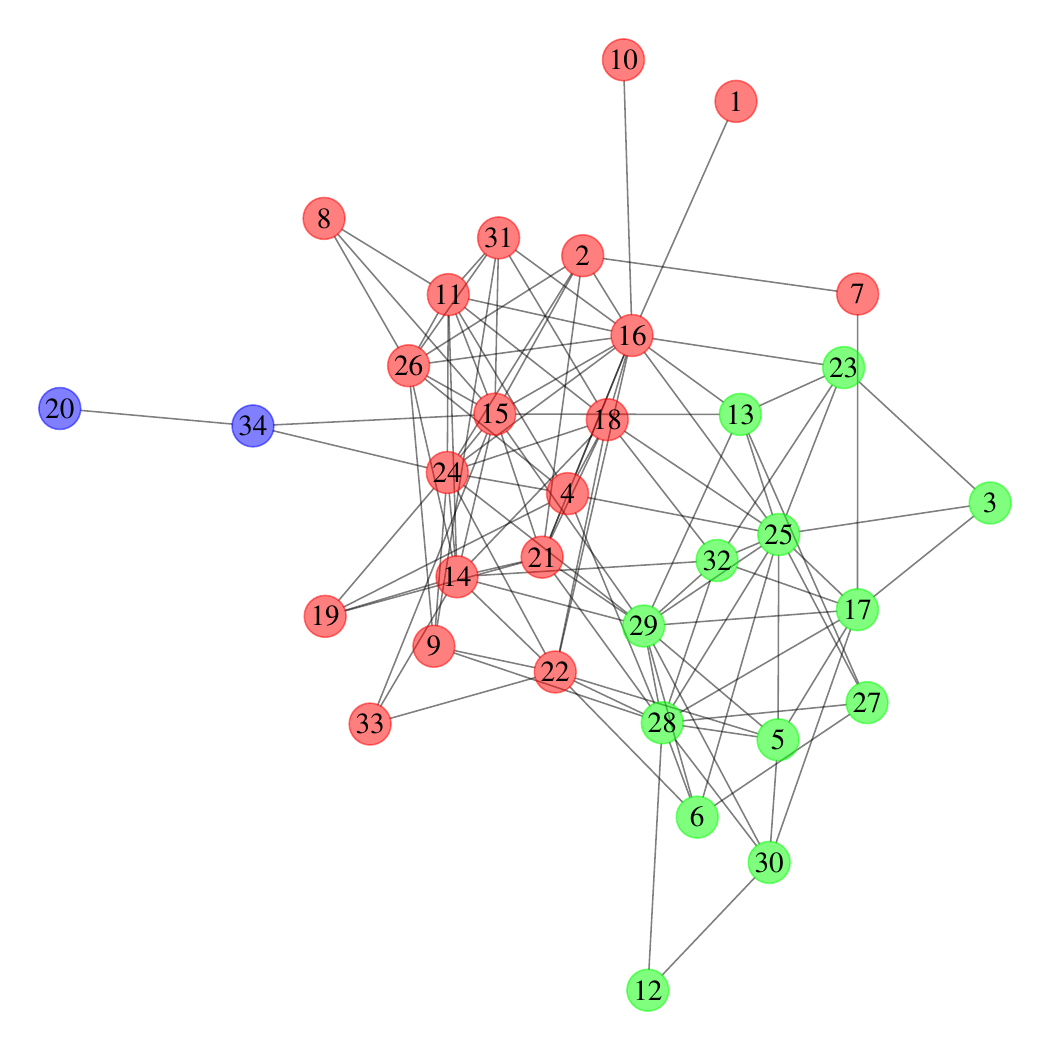}
    }
    \caption{Lazega dataset with different colors representing distinct communities.}
    \label{fig_lazega_net}
\end{figure}

In this section, we analyze the Lazega dataset, which represents collaborative working relationships among members of a New England law firm, as visualized in Figure \ref{fig_lazega_net}. Collected by Lazega \citep{lazega2001collegial}, this undirected binary network was designed to study cooperation and resource exchange among organizational actors. The dataset focuses on a corporate law firm (SG\&R) operating in three offices across Northeastern U.S. cities during the period 1988–1991. It includes data on 71 attorneys (approximately half partners and half associates). However, the analysis presented here is based on a publicly available subset of the data provided in the \texttt{igraph} library, specifically focusing on the 36 partners of the firm. For further details on the broader study and the multiplex nature of these relationships, refer to \cite{lazega1999multiplexity}.

\subsubsection{Exploratory data analysis}

The network analysis reveals 34 nodes and 115 edges, resulting in a sparse structure with a density of 0.205, meaning only about 20\% of possible edges are present. The diameter is 5, with an average path length of 2.14, indicating a compact structure where most nodes are closely connected. A global clustering coefficient of 0.389 suggests moderate clustering, with nodes forming localized groups. The degree assortativity coefficient of -0.168 indicates slight disassortative mixing, where high-degree nodes tend to connect with low-degree nodes.

The average degree is 6.76, indicating each node is connected to around seven others on average. The degree distribution highlights variability, with some nodes substantially more connected. Betweenness centrality identifies key intermediaries, such as actors \(16\) and \(17\), while closeness centrality shows these nodes are efficiently positioned to reach others. Eigenvector centrality highlights actor \(17\) as the most influential node, likely due to its strong connections to well-connected nodes.

Community detection using the fast greedy algorithm (e.g., \citealt{jain1999data}; \citealt{clauset2004finding}) identifies three distinct communities (see panel (b) of Figure \ref{fig_lazega_net}) with a modularity score of 0.313, suggesting a moderate level of community structure.  Nodes within the same community are more densely connected than those in different communities, reflecting a moderately organized network with local clustering and influential nodes. Overall, the network displays a mix of local clustering, influential nodes, and community groupings, reflecting a moderately organized and structured interaction pattern.

\subsubsection{Model fitting}

We implement the sociality model using both MCMC and VI methods (Section \ref{sec_computation}), with hyperparameters set to $a_\sigma = a_\tau = 2$ and $b_\sigma = b_\tau = 1/3$ (Section \ref{sec_prior_elicitation}). Alternative prior specifications are examined as part of the sensitivity analysis detailed in Section \ref{sec_sensitivity_analysis}. The implementation is carried out in R software version 4.4.2, using RStudio version 2024.12.0 Build 467 as the IDE, and executed on a standard laptop equipped with an 11th Gen Intel(R) Core(TM) i7-1165G7 processor (2.80 GHz) and 8.00 GB of RAM. The MCMC algorithm completes in approximately 17 minutes, while the variational approach converges in 0.398 seconds. As discussed below, MCMC captures the full posterior with high precision but is computationally intensive, while VI trades some accuracy for significantly improved computational efficiency.

For the MCMC computation, the results presented below are based on \( B = 25,000 \) posterior samples obtained by thinning the original Markov chains every 10 observations, following a burn-in period of 10,000 iterations. The effective sample sizes are 21,938 for \(\mu\), 25,000 for \(\sigma^2\), and 21,612 for \(\tau^2\), with a mean effective sample size of 24,369 for \(\delta_1, \ldots, \delta_n\). The Monte Carlo standard errors are minimal, ranging from 0.0005 to 0.0039 across the entire set of parameters. These metrics, along with a traceplot of the log-likelihood (not shown here), indicate no evidence of convergence issues.

For computation using VI, the algorithm is initialized randomly, with the mean components set to 0 and the variance components set to 1, perturbed using uniformly distributed random jitter. The shape and rate parameters are initialized based on their corresponding hyperparameters. Convergence is achieved in just 69 iterations, defined as the difference in ELBO values between consecutive iterations being less than \( 10^{-6} \). The algorithm reaches a final ELBO value of -458.4234.

Finally, inference on the model parameters is performed based on the selected fitting approach. MCMC estimates posterior quantities using samples drawn directly from the posterior distribution, ensuring high accuracy in both estimation and uncertainty quantification. In contrast, VI approximates posterior quantities through fitted variational distributions defined by the estimated variational parameters, offering a computationally efficient alternative that balances speed with a reasonable level of approximation accuracy.

\subsubsection{Inference}

\begin{figure}[!htb]
    \centering
    \subfigure[Posterior of $\mu$.]{
        \includegraphics[scale=0.27]{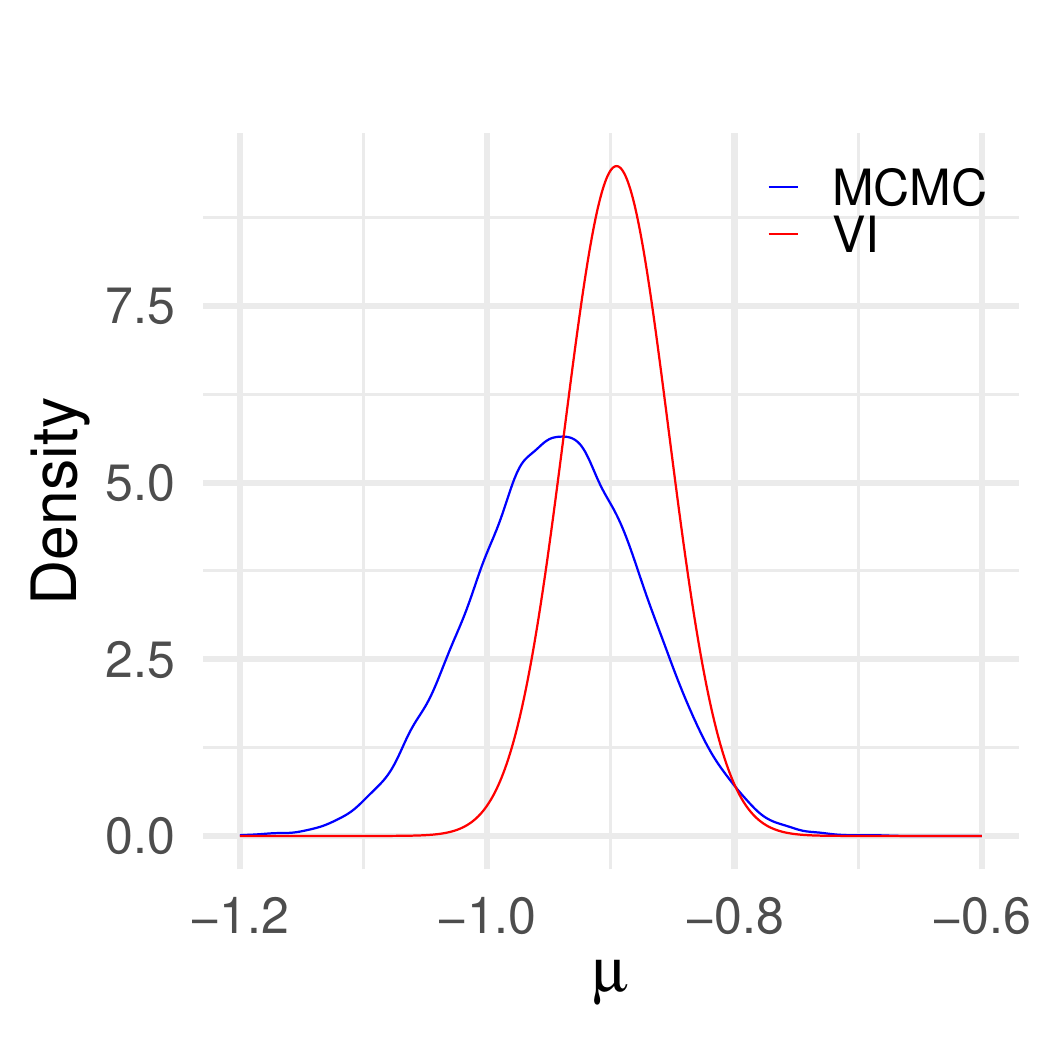}
    }
    \hspace{0em} 
    \subfigure[Posterior of $\sigma^2$.]{
        \includegraphics[scale=0.27]{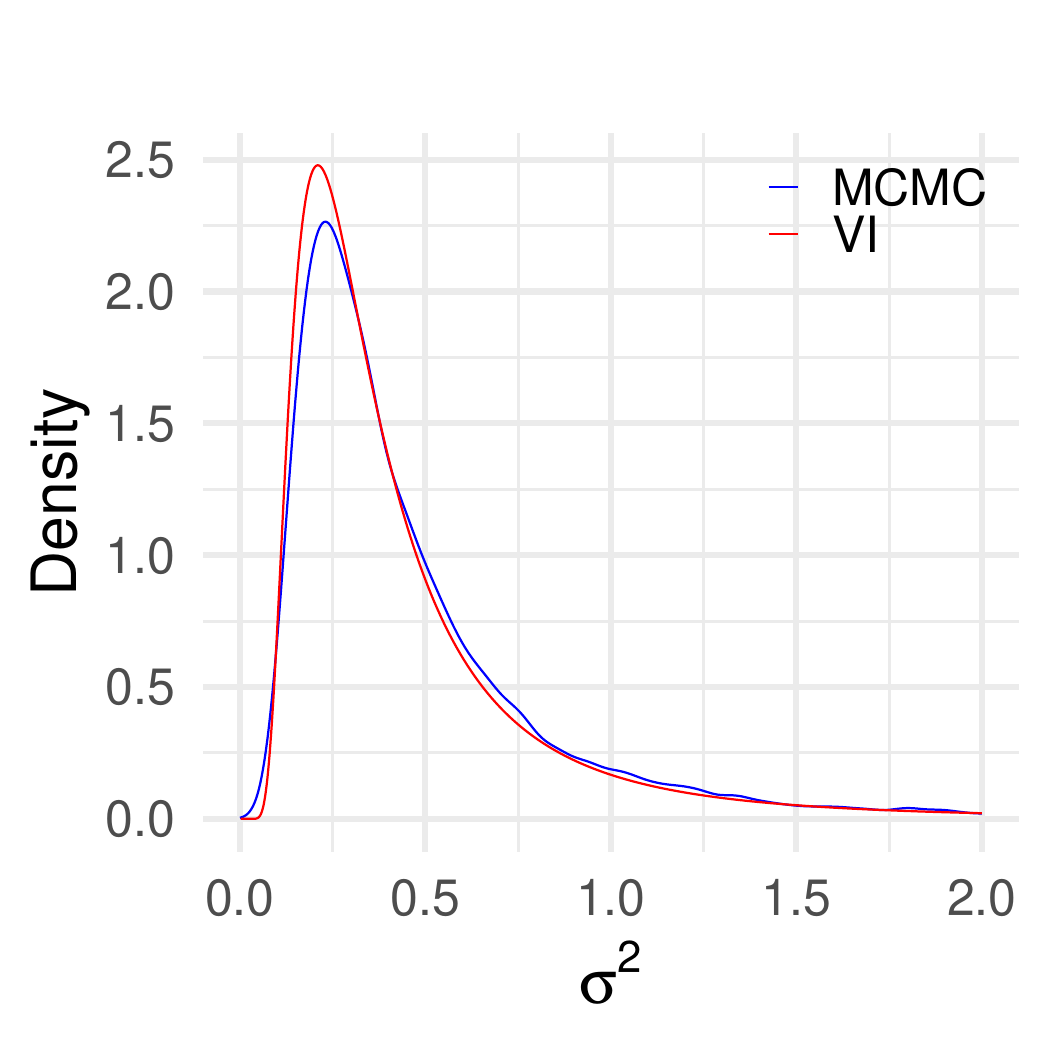}
    }
    \hspace{0em} 
    \subfigure[Posterior of $\tau^2$.]{
        \includegraphics[scale=0.27]{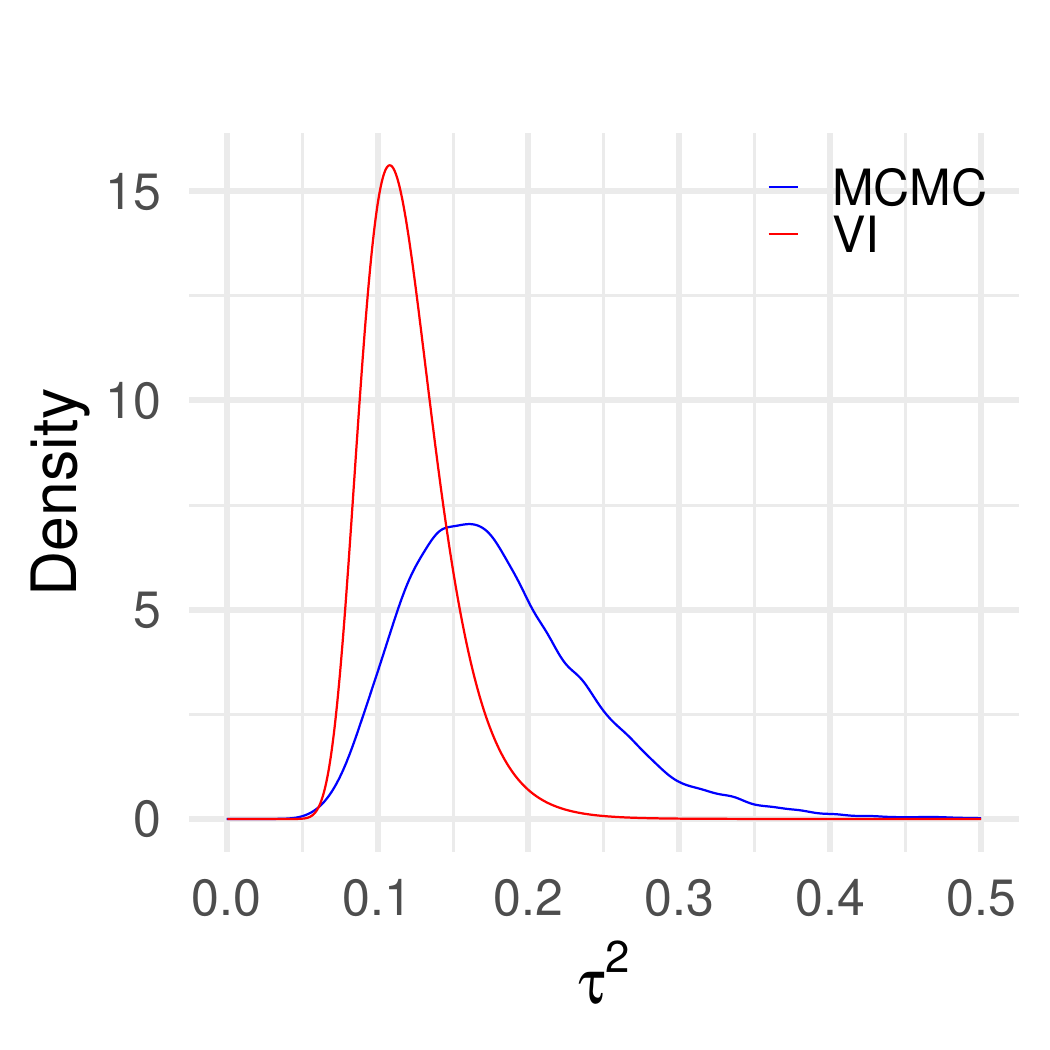}
    }
    \caption{Posterior distributions of $\mu$, $\sigma^2$, and $\tau^2$ estimated using Markov Chain Monte Carlo (MCMC) and Variational Inference (VI).}
    \label{fig_lazega_posterior_mu_sigma2_tau2}
\end{figure}

Figure \ref{fig_lazega_posterior_mu_sigma2_tau2} and Table \ref{tab_lazega_posterior_mu_sigma2_tau2} summarize the posterior distributions of $\mu$, $\sigma^2$, and $\tau^2$ estimated using MCMC and VI. Based on the MCMC results, the negative mean of $\mu$ indicates a low baseline probability of forming ties, reflecting the competitive and hierarchical nature of law firms. The narrow credible interval underscores the selective and uneven distribution of collaborations among partners.

\begin{table}[!htb]
\centering
\begin{tabular}{ccccc}
\hline
\multicolumn{5}{c}{Markov chain Monte Carlo} \\
\hline
Parameter & Mean & SD & $\text{Q}_{2.5\%}$ & $\text{Q}_{97.5\%}$ \\
\hline
$\mu$       & -0.9438  & 0.0696 & -1.0810  & -0.8093 \\
$\sigma^2$  & 0.5211   & 0.6224 &  0.1201  &  1.9308 \\
$\tau^2$    & 0.1806   & 0.0636 &  0.0867  &  0.3333 \\
\hline
\multicolumn{5}{c}{Variational Inference} \\
\hline
Parameter & Mean & SD & $\text{Q}_{2.5\%}$ & $\text{Q}_{97.5\%}$ \\
\hline
$\mu$       & -0.8954  & 0.0421 & -0.9779  & -0.8129 \\
$\sigma^2$  &  0.4901  & 0.6930 &  0.1146  &  1.7687 \\
$\tau^2$    &  0.1201  & 0.0291 &  0.0760  &  0.1890 \\
\hline
\end{tabular}
\caption{Posterior summary statistics for $\mu$, $\sigma^2$, and $\tau^2$.}
\label{tab_lazega_posterior_mu_sigma2_tau2}
\end{table}

The estimate of the variance component $\sigma^2$ highlights fluctuations in tie formation. The wide credible interval suggests heterogeneity driven by structural factors such as office locations or practice areas. In contrast, the small mean of the variance component $\tau^2$ indicates limited variation in individual deviations from the global baseline connectivity, reflecting the structured and formalized nature of corporate law firm partnerships shaped by professional norms and organizational frameworks.

The greater variability in global connectivity compared to individual sociality effects suggests that tie formation is primarily influenced by network-level factors rather than individual behaviors. Structural elements such as inter-office dynamics, practice group alignment, and firm-wide policies play a pivotal role, emphasizing the hierarchical nature of collaboration within the firm. Incorporating covariates such as office location, seniority, or practice specialization could provide deeper insights into the observed variability in tie formation.

While the posterior means from MCMC and VI are generally consistent, VI often underestimates uncertainty for $\mu$ and $\tau^2$ and occasionally overestimates it for $\sigma^2$. These differences highlight the trade-off between the computational efficiency of VI and the more reliable uncertainty quantification provided by MCMC. For applications where accurate uncertainty estimates are critical, MCMC is the preferred method, whereas VI serves as a faster, approximate alternative when computational efficiency is prioritized.

\begin{figure}[!htb]
    \centering
    \subfigure[MCMC.]{
        \includegraphics[scale=0.42]{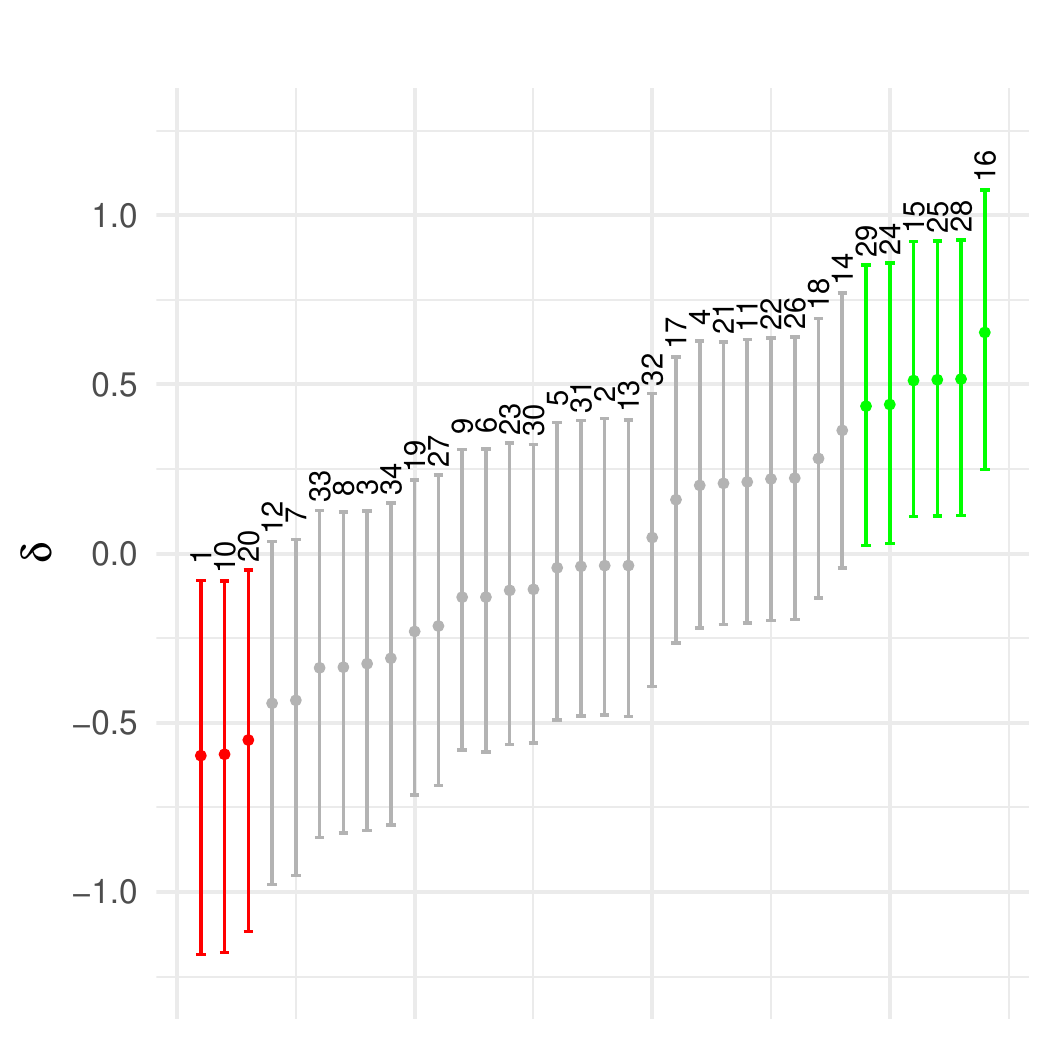}
    }
    \hspace{0em} 
    \subfigure[VI.]{
        \includegraphics[scale=0.42]{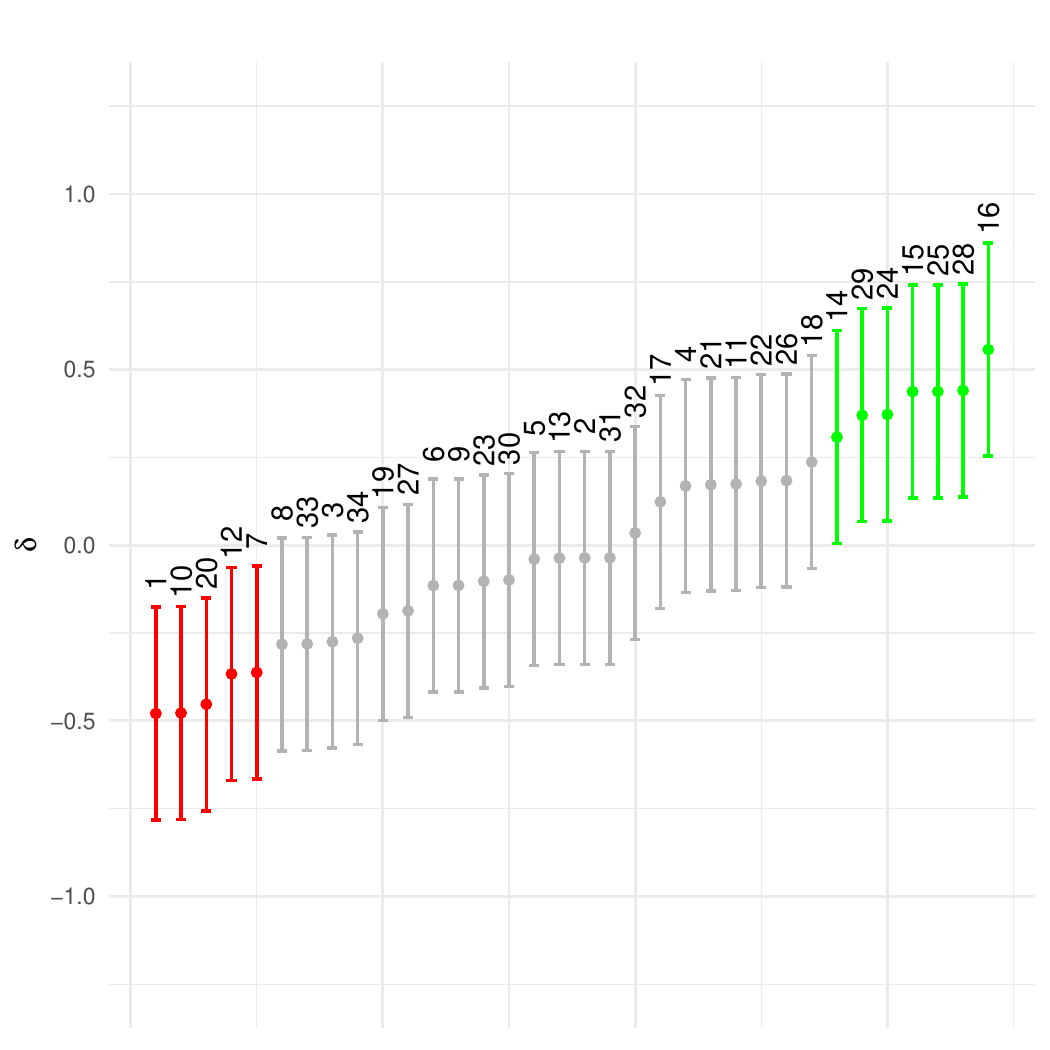}
    }
    \caption{Posterior means and 95\% credible intervals for \(\delta_1, \ldots, \delta_n\) estimated using Markov Chain Monte Carlo (MCMC) and Variational Inference (VI). The interval colors indicate whether the effects are significantly less than zero (red), include zero (gray), or are significantly greater than zero (green).}
    \label{fig_lazega_posterior_delta}
\end{figure}

Figure \ref{fig_lazega_posterior_delta} displays the posterior means and 95\% credible intervals for the sociality effects \(\delta_1, \ldots, \delta_n\) estimated using MCMC and VI. The interval colors indicate whether the effects are significantly less than zero (red), include zero (gray), or are significantly greater than zero (green). Based on the MCMC results, actors classified as red (significantly less than zero) include 3, 6, 9, 13, 23, 30, 33, and 34, accounting for 22.22\% of the total. Actors with gray intervals (including zero) are 1, 4, 7, 8, 10, 11, 12, 14, 17, 18, 19, 21, 22, 24, 26, 27, 29, 31, and 32, representing 52.78\%. Lastly, actors classified as green (significantly greater than zero) are 2, 5, 15, 16, 25, and 28, comprising 16.67\% of the total.

Actors with red intervals exhibit sociality effects credibly lower than the global baseline, indicating reduced involvement in forming connections within the network. This behavior may reflect peripheral or specialized roles within the firm, where certain partners primarily interact within specific subgroups or have responsibilities confined to particular areas. In contrast, green intervals correspond to actors with significantly higher effects, reflecting greater participation in the collaborative network. These partners likely serve as key figures, such as leaders or connectors between offices or practice groups, whose central roles enhance overall network connectivity. Gray intervals, which include zero, suggest that the sociality effects of these actors are not significantly different from the baseline, representing average participation in the network. Together, these findings underscore the hierarchical and structured nature of the law firm, where collaborations are unevenly distributed and influenced by factors such as location, specialization, and seniority.

The results based on VI exhibit a similar overall pattern to those generated using MCMC, with the same actors classified as red, gray, and green. However, the credible intervals produced by VI are narrower, reflecting its tendency to underestimate uncertainty. This narrowing can potentially lead to misclassifications, where certain actors having null sociality effects (red or green) by VI might instead fall into the gray category when considering the broader intervals provided by MCMC. For instance, some partners with non-null sociality effects in VI may be borderline cases, with their true classification appearing more uncertain under the more robust uncertainty estimates from MCMC.

The differences between MCMC and VI carry important implications. While VI offers greater computational efficiency, its tendency to underestimate uncertainty can lead to less reliable interpretations, particularly in scenarios where precise credible intervals and statistical significance are essential for decision-making. In contrast, MCMC provides more robust and reliable estimates, enabling a deeper and more accurate interpretation of collaboration patterns within the firm. The MCMC results reveal a structured and hierarchical collaborative network, where partner roles are shaped by both global and individual factors. Incorporating covariates such as office location, seniority, or specialization could provide greater insight into the factors driving the observed sociality effects.


Estimating sociality effects is particularly useful as it facilitates clustering tasks to segment individuals within the network. Using MCMC, clustering can be performed by grouping sociality effects at each iteration with standard methods like k-means (e.g., \citealt{james2023introduction}), producing cluster indicators \(\xi_1, \ldots, \xi_n\), where \(\xi_i = k\) denotes that individual \(i\) belongs to cluster \(k\). By obtaining the indicator variables \(\xi_i^{(b)}\) at each iteration, for \(i = 1, \ldots, n\) and \(b = 1, \ldots, B\), the posterior probability that individuals \(i\) and \(j\) belong to the same cluster can be approximated as  
\[
\textsf{P}(\xi_i = \xi_j \mid \mathbf{Y}) \approx \frac{1}{B}\sum_{b=1}^B I(\xi_i^{(b)} = \xi_j^{(b)}),
\]
where \(I(\cdot)\) is the indicator function. These coclustering probabilities serve as a foundation for posterior inference regarding the partitioning of actors within the network.

\begin{figure}[!htb]
    \centering
    \subfigure[Coclustering probabilities.]{
        \includegraphics[scale=0.42]{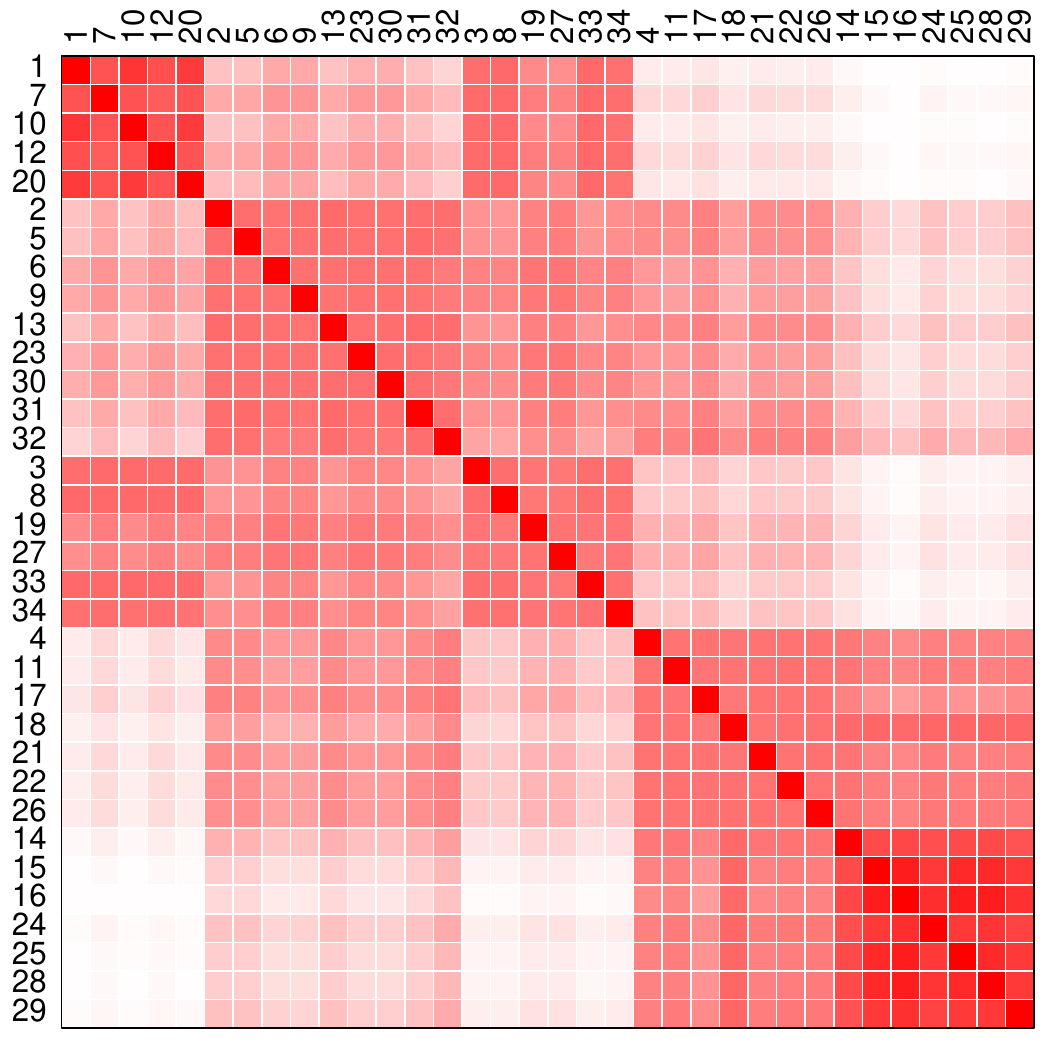}
    }
    \hspace{0em} 
    \subfigure[Partition.]{
        \includegraphics[scale=0.42]{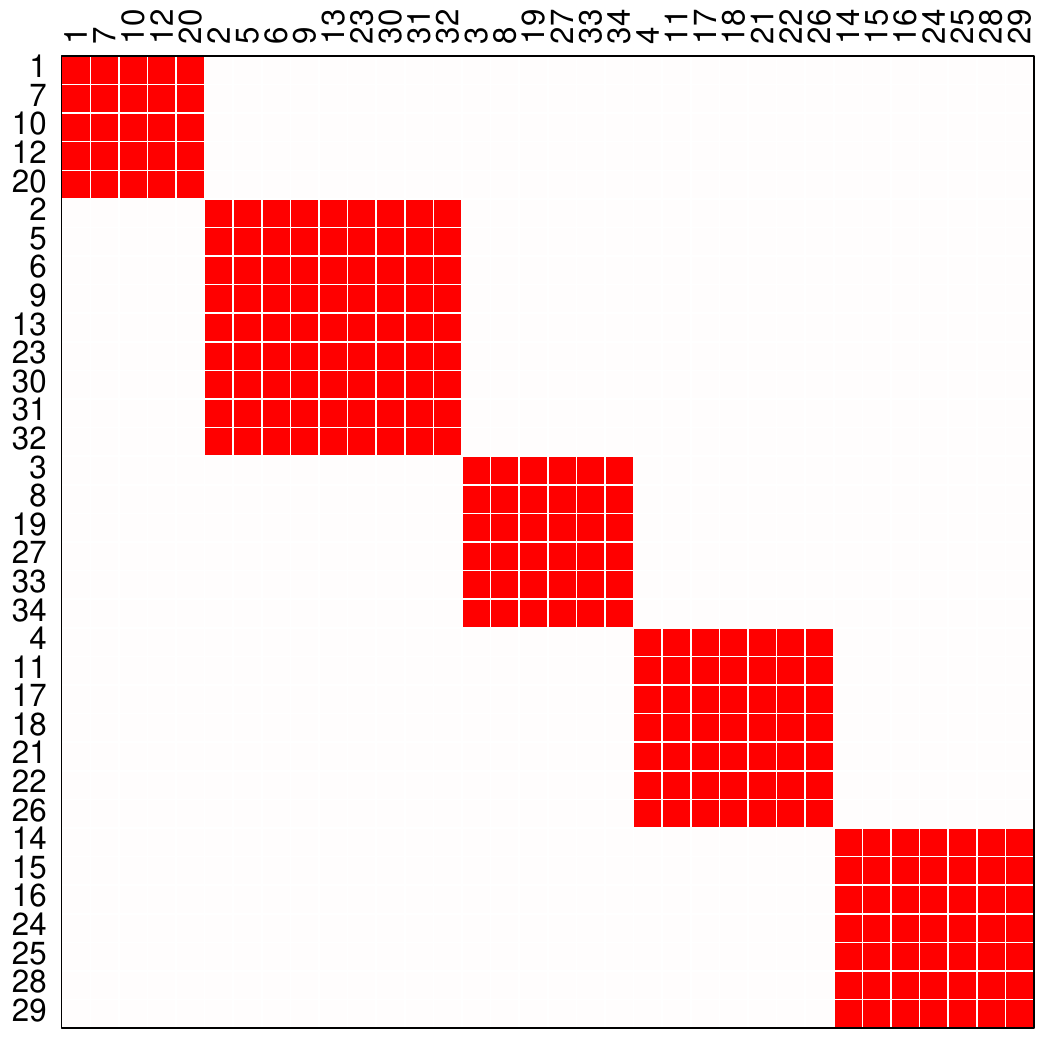}
    }
    \caption{Clustering network actors using posterior samples of the sociality effects.}
    \label{fig_lazega_posterior_clustering}
\end{figure}

Panel (a) of Figure \ref{fig_lazega_posterior_clustering} presents the coclustering probability matrix, computed using k-means with the elbow method to determine the optimal number of clusters at each iteration. Darker cells represent higher probabilities, indicating stronger pairwise affinities. This matrix captures the posterior probabilities that two actors belong to the same cluster based on their sociality effects. Panel (b) shows a point estimate of the partition derived from these probabilities, determined using BIC through mixture model estimation \citep{scrucca2016mclust}, with matrices reordered by cluster membership for clarity. The partition reveals well-defined clusters that reflect collaboration patterns within the law firm, grouping partners with similar sociality effects, from highly connected senior members to more peripheral actors.

\begin{figure}[!htb]
    \centering
    \subfigure[Clustering inference on \(\boldsymbol\delta\) using MCMC.]{
        \includegraphics[scale=0.42]{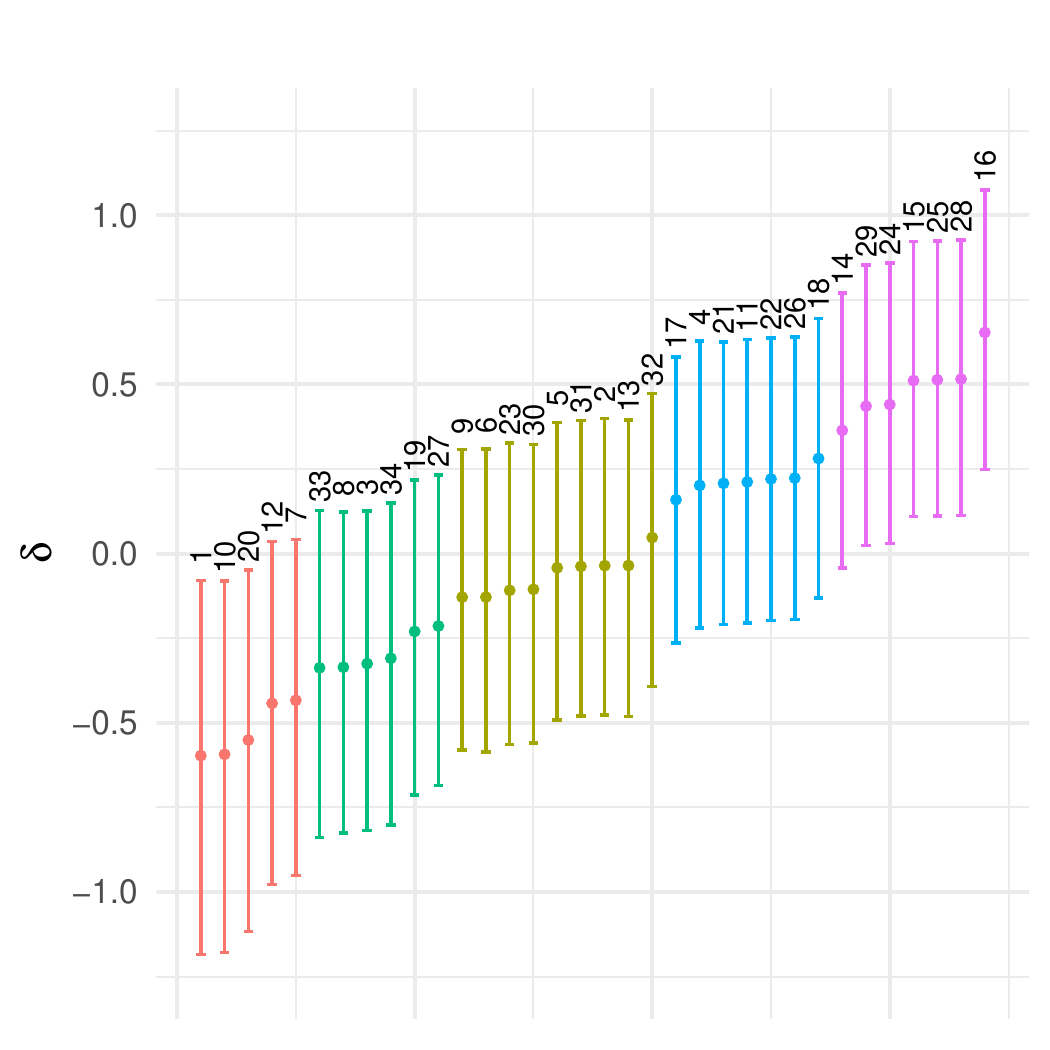}
    }
    \hspace{0em} 
    \subfigure[Clustering inference on \(\boldsymbol\delta\) using VI.]{
        \includegraphics[scale=0.42]{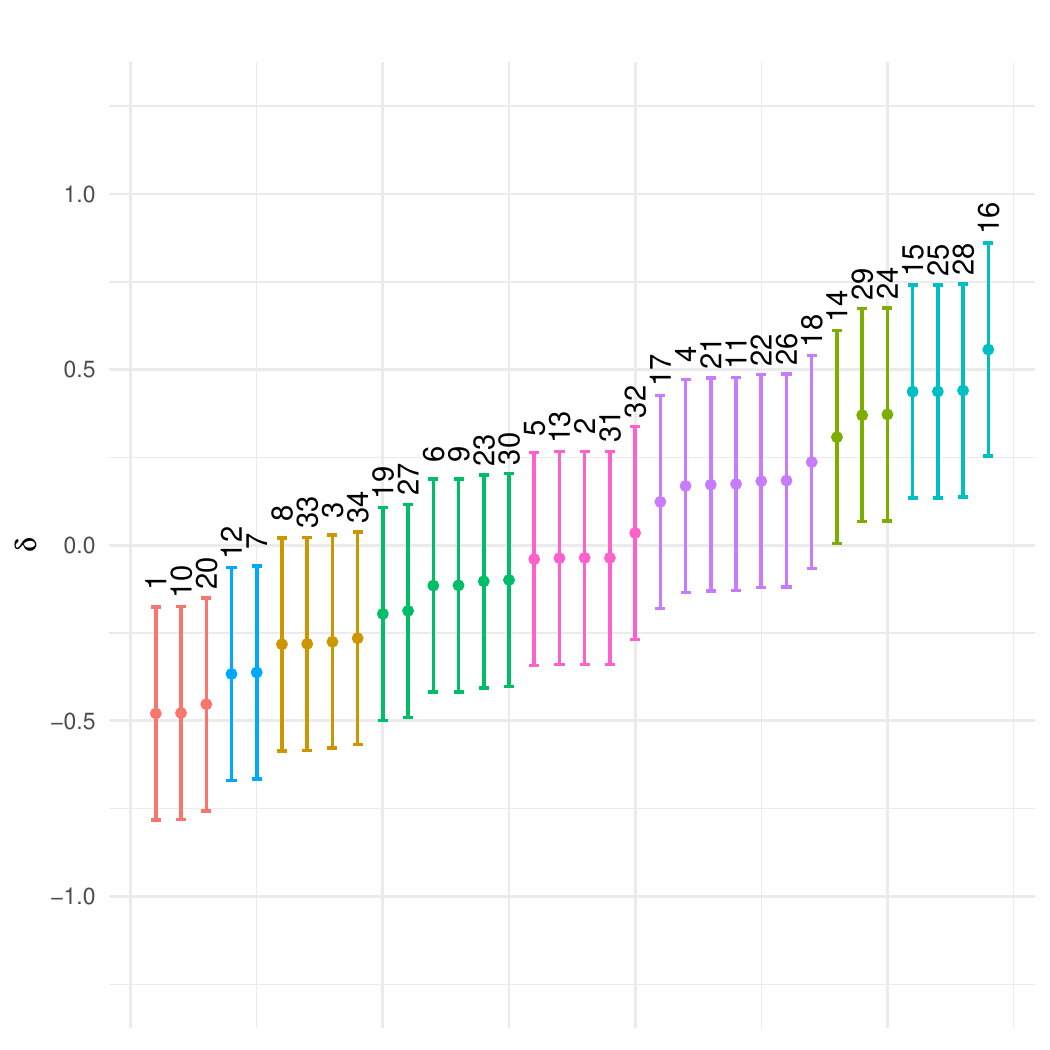}
    }
    \caption{Clustering network actors based on their individual sociality effects.}
    \label{fig_lazega_posterior_delta_clustering}
\end{figure}

Using VI, a partition can be obtained by directly applying the chosen clustering method to the variational posterior means of the sociality effects. Figure \ref{fig_lazega_posterior_delta_clustering} presents the point estimates along with 95\% credible intervals grouped based on their sociality effects according to the methods described above using MCMC and VI. MCMC-based clustering groups actors according to their sociality effects, capturing a spectrum from peripheral individuals to highly connected ones, resulting in 5 clusters. VI-based clustering, while exhibiting similar overall patterns, produces 8 tighter clusters, offering finer segmentation. This consistency in broad clustering patterns reflects the structured nature of the firm’s collaboration network. However, MCMC adopts a more conservative and robust approach by accounting for variability in sociality effects, while VI prioritizes computational efficiency, generating sharper but potentially less reliable clusters.

\subsubsection{Sensitivity analysis}\label{sec_sensitivity_analysis}

Now, we conduct a sensitivity analysis on the choice of the model's hyperparameters. To this end, we fit the model using both MCMC and VI under the same settings as in the data analysis, considering six different prior configurations, namely, Prior 1: \( a_\sigma = a_\tau = 2 \) and \( b_\sigma = b_\tau = 1/3 \); Prior 2: \( a_\sigma = a_\tau = 3 \) and \( b_\sigma = b_\tau = 1/3 \); Prior 3: \( a_\sigma = a_\tau = 2 \), \( b_\sigma = 1/2 \), and \( b_\tau = 1/4 \); Prior 4: \( a_\sigma = a_\tau = 3 \), \( b_\sigma = 1/2 \), and \( b_\tau = 1/4 \); Prior 5: \( a_\sigma = a_\tau = 2 \) and \( b_\sigma = b_\tau = 1 \); and Prior 6: \( a_\sigma = a_\tau = 3 \) and \( b_\sigma = b_\tau = 2 \).

\begin{figure}[!htb]
    \centering
    \includegraphics[scale=0.5]{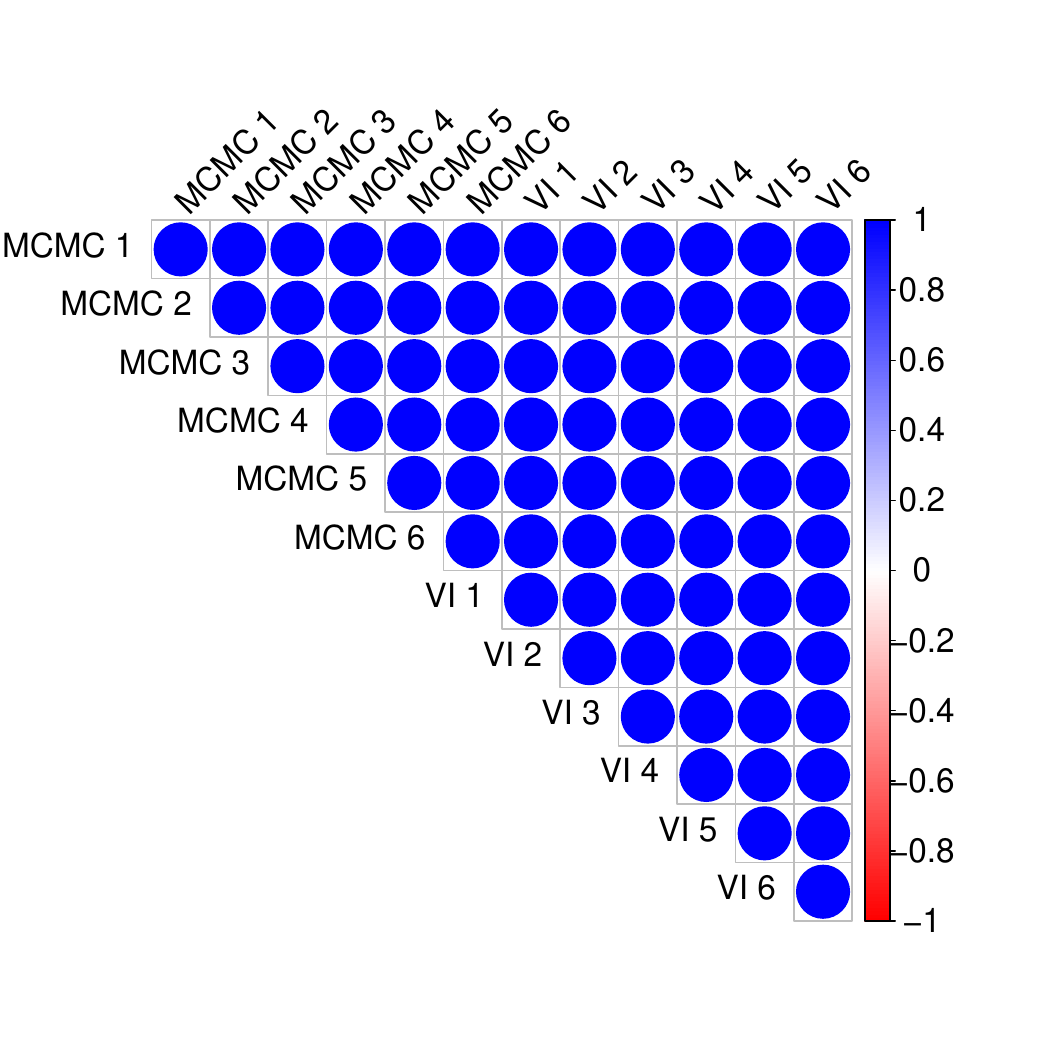}
    \caption{Correlation matrix visualization of sociability parameter estimates under six different prior configurations, comparing MCMC and VI approaches.}
    \label{fig_lazega_sensitivity}
\end{figure}

For each case, we obtain estimates of each subject-specific sociability parameter \( \delta_i \) and compute a correlation matrix to assess the similarity of the estimates across configurations. Figure \ref{fig_lazega_sensitivity} presents a visualization of this correlation matrix, showing that the sociability parameter estimates remain highly consistent regardless of the estimation method or prior configuration. This result indicates that the sociability model is highly robust to both prior specification and estimation approach.

\subsubsection{Model comparison via posterior predictive distributions}

From now on, we compare the sociality model with the latent space models described in Section \ref{sec_latent_space_models}. These models are fitted using MCMC with $B = 25,000$ posterior samples, obtained by thinning the original Markov chains every 10 iterations after a burn-in period of 10,000 iterations. In particular, the latent space models are fitted using $K=4$ for the distance and eigen models, and $K=10$ for the class model. These values have been shown to provide a good balance between goodness-of-fit and model complexity. We refer the reader to \cite{sosa2021review} for a comprehensive review of the latent modeling approach, including detailed explanations of MCMC methods.

\begin{figure}[!htb]
    \centering
    \subfigure[Density.]{
        \includegraphics[scale=0.27]{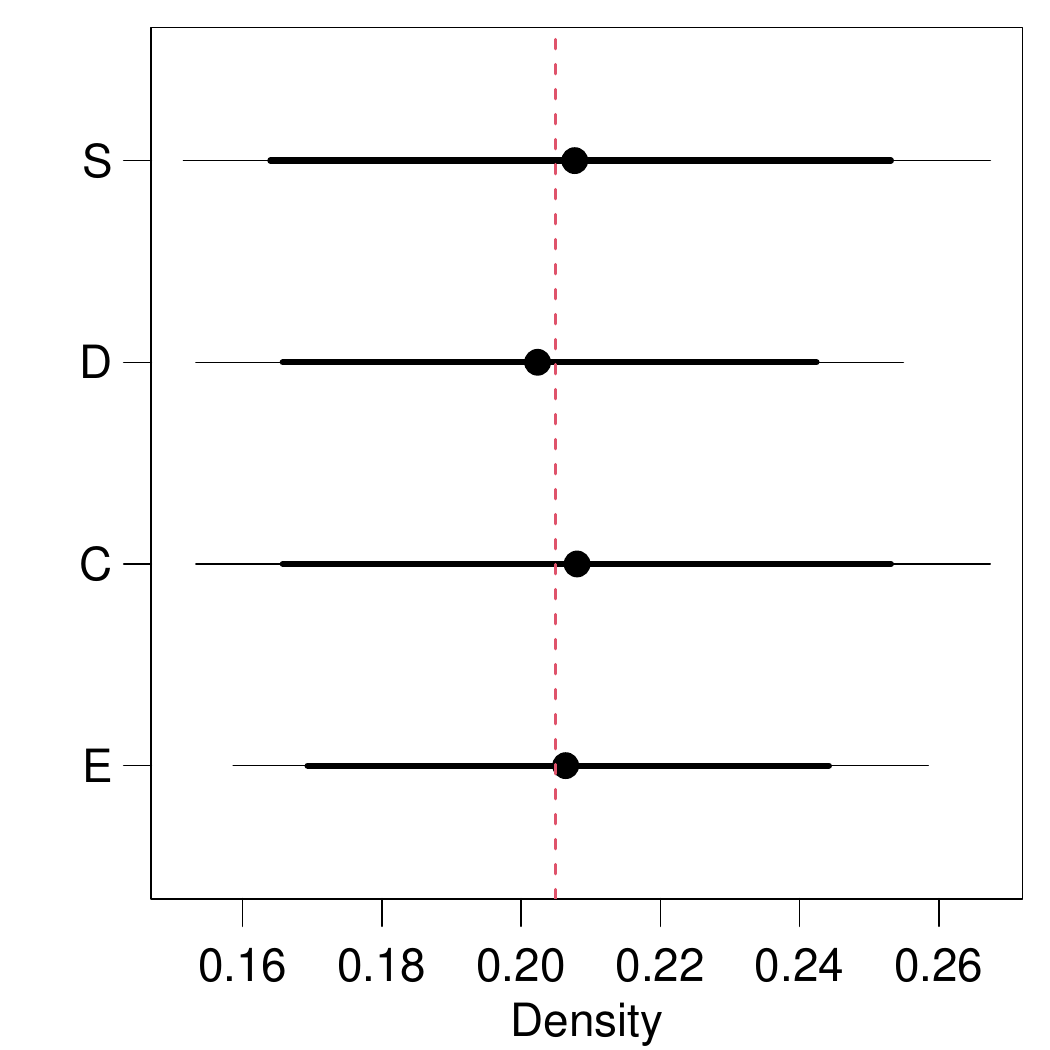}
    }
    \hspace{0em} 
    \subfigure[Transitivity.]{
        \includegraphics[scale=0.27]{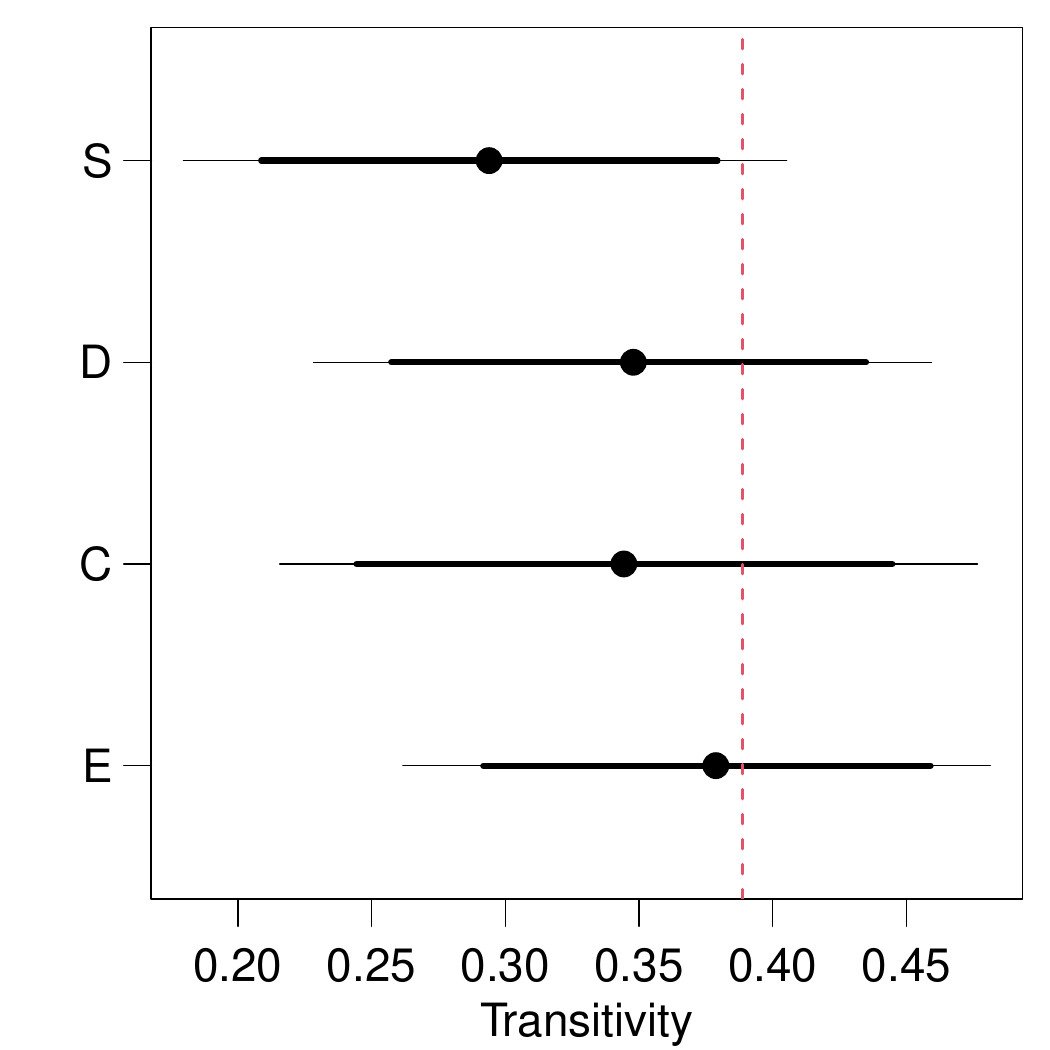}
    }
    \hspace{0em} 
    \subfigure[Degree assortativity.]{
        \includegraphics[scale=0.27]{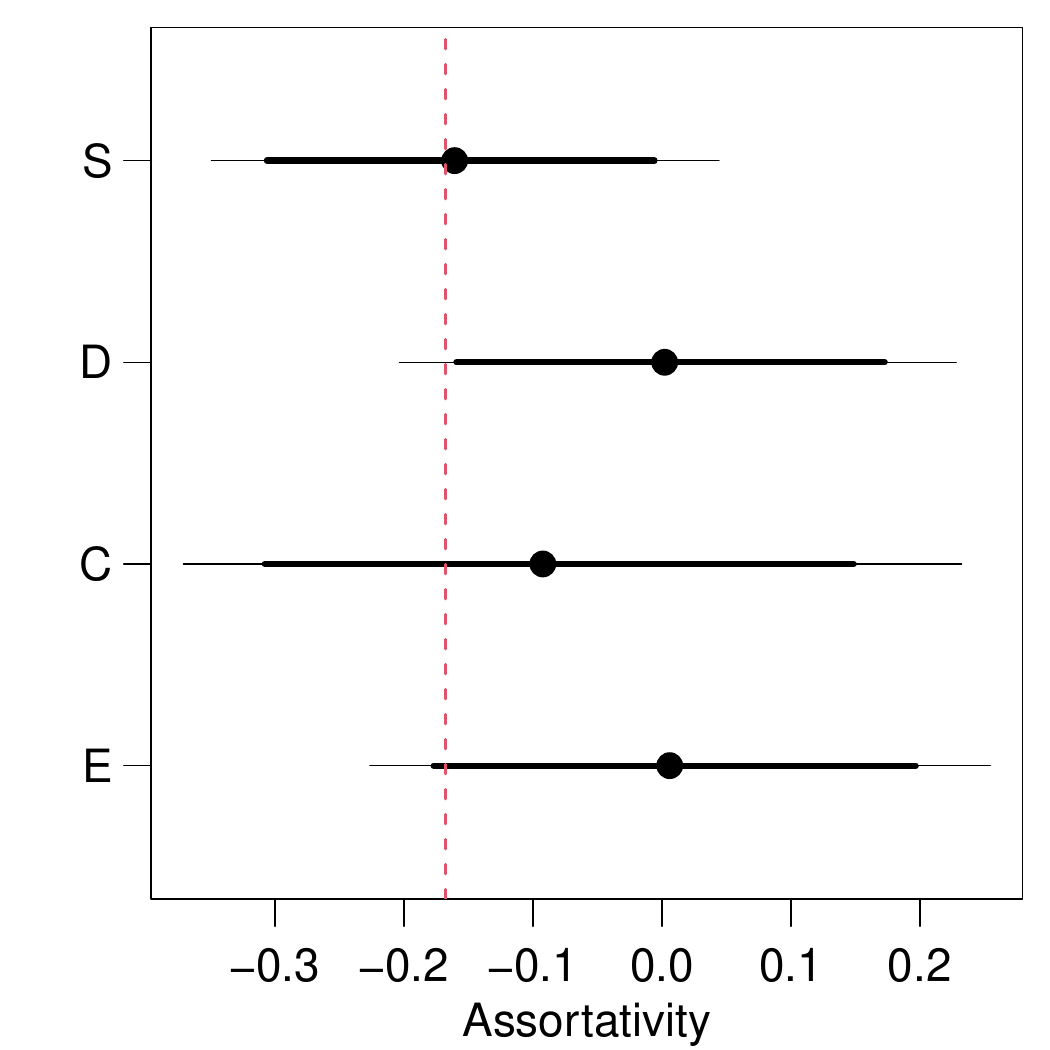}
    }
    \subfigure[Mean geodesic distance.]{
        \includegraphics[scale=0.27]{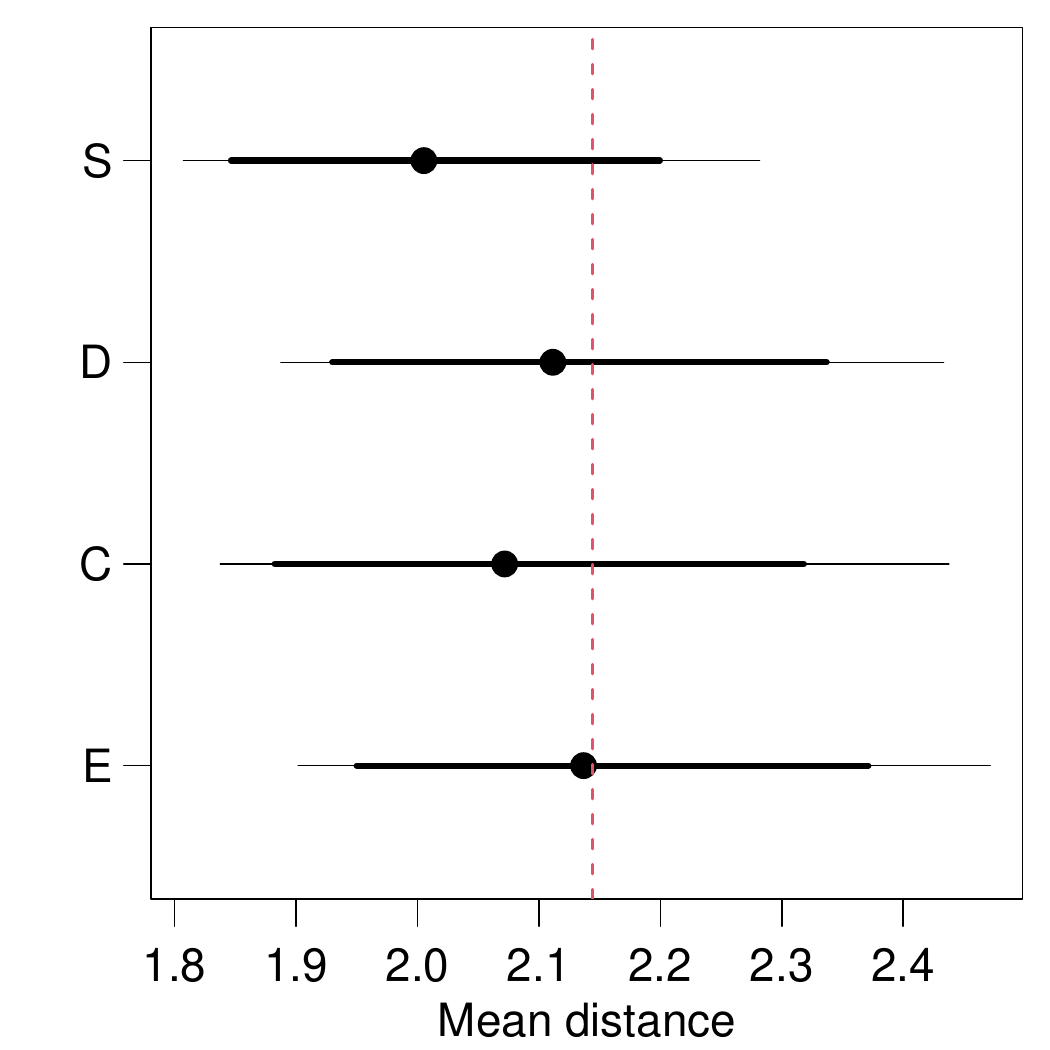}
    }
    \hspace{0em} 
    \subfigure[Mean degree.]{
        \includegraphics[scale=0.27]{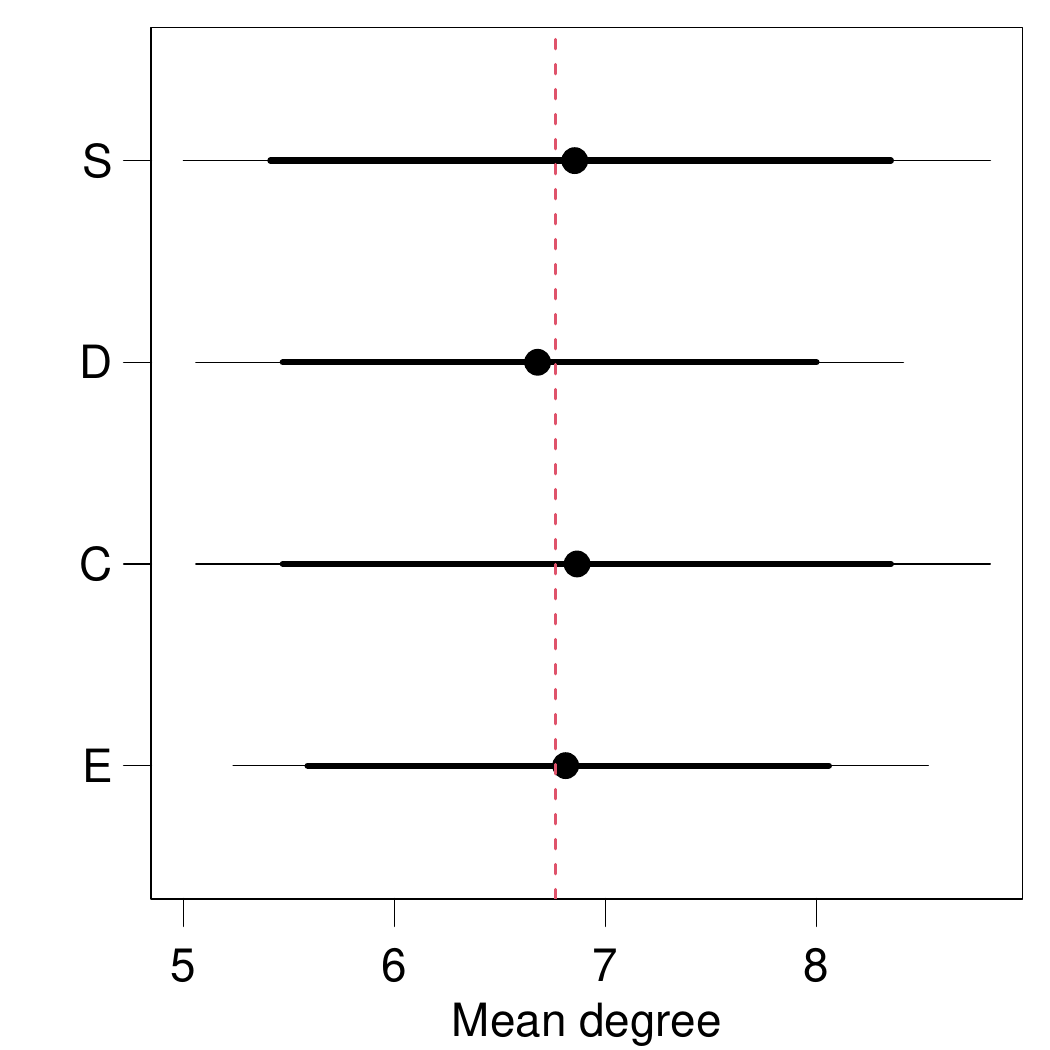}
    }
    \hspace{0em} 
    \subfigure[SD degree.]{
        \includegraphics[scale=0.27]{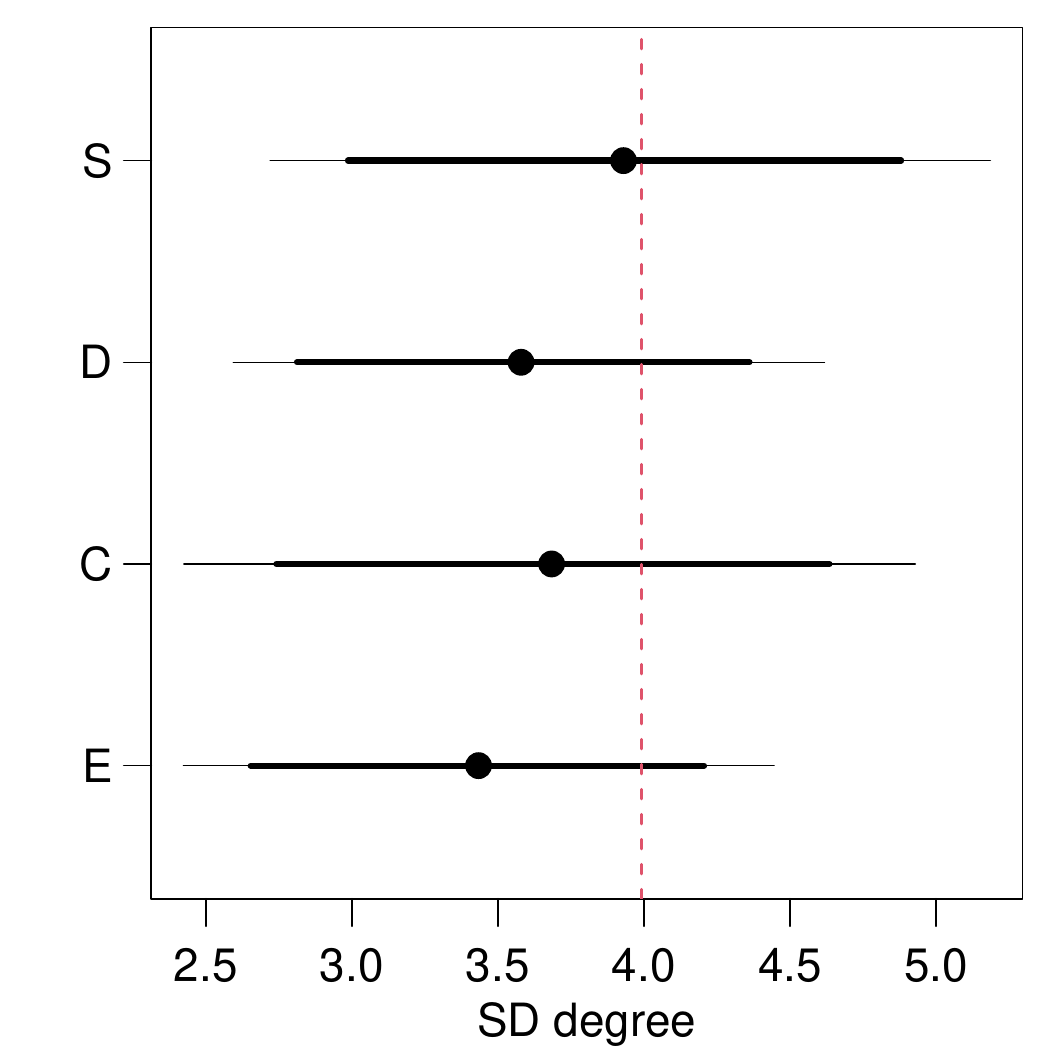}
    }
    \caption{Posterior mean (black circle) with 95\% (thin line) and 99\% (thick line) credible intervals for the empirical distribution of test statistics from replicated data, alongside the observed value (red dashed line), comparing the sociality (S), distance (D), class (C), and eigen (E) models using the Lazega dataset.}
    \label{fig_lazega_test_statistics}
\end{figure}

In this way, following the approach of \cite[Chap. 6]{gelman2014bayesian} and \cite[Chap. 4]{kolaczyk2020statistical}, we generate pseudo-data from all fitted models and compute a set of summary statistics (density, transitivity, assortativity, mean geodesic distance, mean degree, and standard deviation of degree) for each posterior sample across the selected networks. This allows us to estimate the posterior predictive distribution of these summaries, which we then compare to the observed values in the original dataset (Figure \ref{fig_lazega_test_statistics}).

All models adequately reproduce the evaluated structural characteristics, though some important differences emerge. The sociality model tends to underestimate transitivity while capturing degree assortativity more accurately than the other models, which tend to overestimate it. Similarly, it provides the best fit for the standard deviation of degree, whereas the other models tend to underestimate it. This behavior is expected, as the sociality model is formulated exclusively in terms of actors' popularity. Overall, we conclude that the Sociality model is competitive compared to other models in the literature.

\subsubsection{Model fit}

Continuing with the model comparison, to assess the goodness-of-fit of each model, we compare them using criteria that account for both model fit and complexity, serving as essential tools for model selection. Relying solely on the highest log-likelihood for model selection is generally not advisable, as it can lead to overfitting. A robust selection process must consider model complexity by incorporating the number of estimated parameters, ensuring an optimal balance between fit and parsimony. To address this, several metrics have been developed to evaluate both goodness-of-fit and in-sample predictive performance. Among these, the Deviance Information Criterion (DIC; see \citealt{spiegelhalter2002bayesian, spiegelhalter2014deviance}) and the frequently used methods Applicable Information Criterion (WAIC; see \citealt{watanabe2010asymptotic, watanabe2013widely}) are two widely used methods. A detailed methodological discussion of these criteria is provided in \cite{gelman2014understanding}.

The WAIC is a fully Bayesian approach for evaluating in-sample predictive accuracy based on the posterior predictive distribution. It is defined as:
\[
\text{WAIC} = -2\,\text{lppd} + 2p_{\text{WAIC}},
\]
where \(\text{lppd}\) represents the log pointwise predictive density and is calculated as:
\[
\text{lppd} = \sum_{i=1}^n \log \int_\Theta p(y_i \mid \boldsymbol{\theta}) p(\boldsymbol{\theta} \mid \boldsymbol{y}) \, \mathrm{d}\boldsymbol{\theta} \approx \sum_{i=1}^n \log\left(\tfrac{1}{B} \textstyle\sum_{b=1}^B p(y_i \mid \boldsymbol{\theta}^{(b)})\right).
\]
The effective number of parameters, \(p_{\text{WAIC}}\), is expressed as:
\[
p_{\text{WAIC}} = 2 \sum_{i=1}^n \left(\log \mathsf{E}(p(y_i \mid \boldsymbol{\theta}) \mid \boldsymbol{y}) - \mathsf{E}(\log p(y_i \mid \boldsymbol{\theta}) \mid \boldsymbol{y})\right),
\]
and can be approximated as:
\[
p_{\text{WAIC}} \approx 2 \sum_{i=1}^n \left(\log\left(\tfrac{1}{B} \textstyle\sum_{b=1}^B p(y_i \mid \boldsymbol{\theta}^{(b)})\right) - \tfrac{1}{B} \textstyle\sum_{b=1}^B \log p(y_i \mid \boldsymbol{\theta}^{(b)})\right).
\]
WAIC measures predictive accuracy within a fully Bayesian framework, offering advantages over criteria like DIC by more directly leveraging the posterior predictive distribution. Lower WAIC values indicate a preferred model.

The sociality model has the worst fit (WAIC = 515.67) but is the most parsimonious ($p_{\text{WAIC}} = 21.35$), suggesting that its simplicity limits its ability to fully capture the data structure for this dataset. In contrast, the distance model achieves the best fit (WAIC = 420.54) but is also quite complex ($p_{\text{WAIC}} = 53.49$), indicating that its flexibility enhances its performance. The class and eigen models fall in between, with WAIC values of 465.60 and 421.66 and complexities of 39.67 and 56.73 effective parameters, respectively. While the class model balances fit and parsimony, the eigen model performs closer to the distance model but with greater complexity. These patterns are also observed when evaluating out-of-sample predictive performance, where the sociality model performs almost as well as the other models while remaining the least complex.

\subsubsection{Predictive accuracy}

Now, we perform a 5-fold cross-validation experiment for each model as follows: First, the data is randomly partitioned into five subsets of approximately equal size. For each subset \(s\), we fit the model using only the observed links \(\{y_{i,j}:(i,j)\notin s\}\) and compute \(\textsf{E}(y_{k,\ell} \mid \{y_{i,j}:(i,j)\notin s\})\), the posterior predictive mean for each \(y_{k,\ell}\) in \(s\). These posterior predictive means are then used as inputs to a binary classifier to construct the receiver operating characteristic (ROC) curve. Finally, we assess predictive performance using the area under the curve (AUC), which quantifies the model’s ability to predict missing links. A higher AUC indicates better predictive performance, meaning the model more effectively distinguishes between observed links (1s) and non-links (0s).

\begin{figure}[!b]
    \centering
    \subfigure[Sociality model.]{
        \includegraphics[scale=0.3]{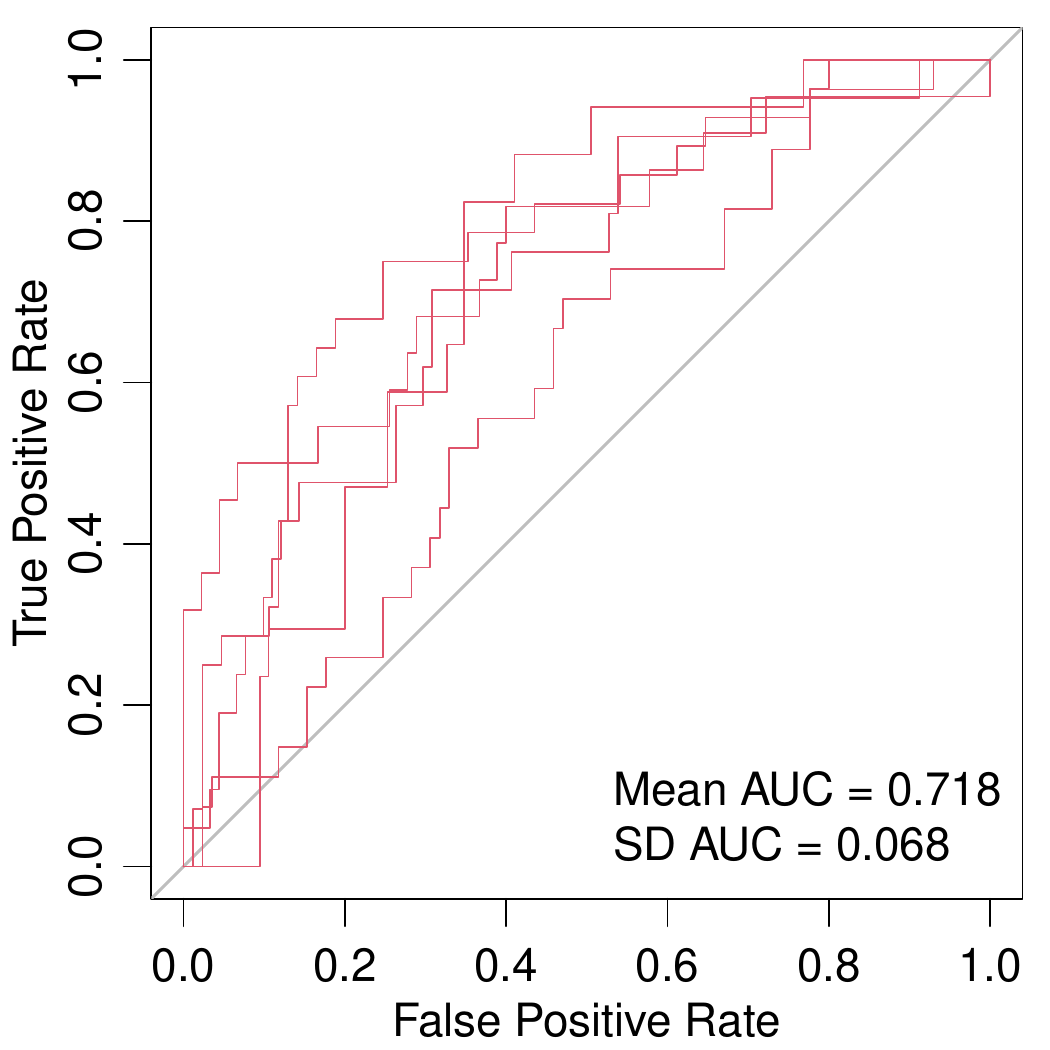}
    }
    \hspace{0em} 
    \subfigure[Distance model.]{
        \includegraphics[scale=0.3]{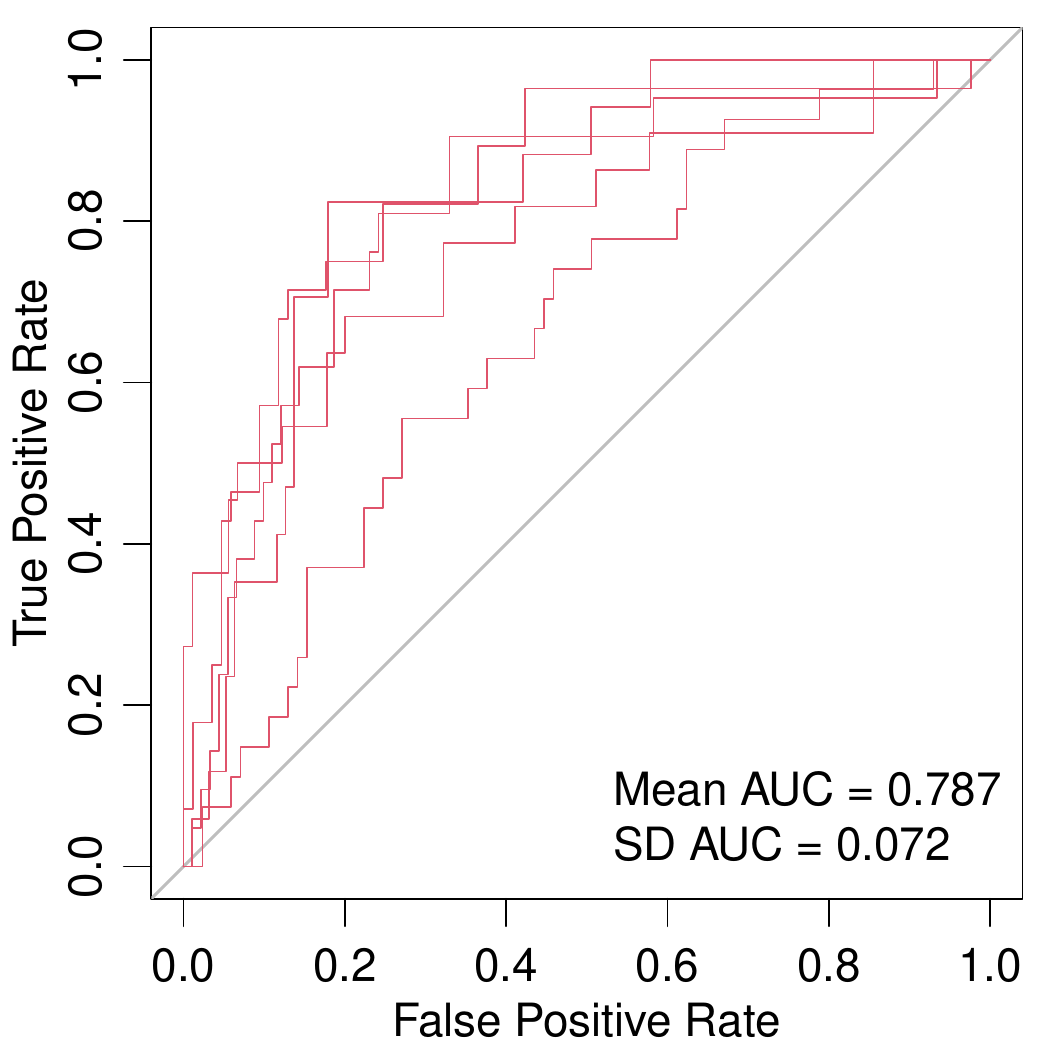}
    }
    \hspace{0em} 
    \subfigure[Class model.]{
        \includegraphics[scale=0.3]{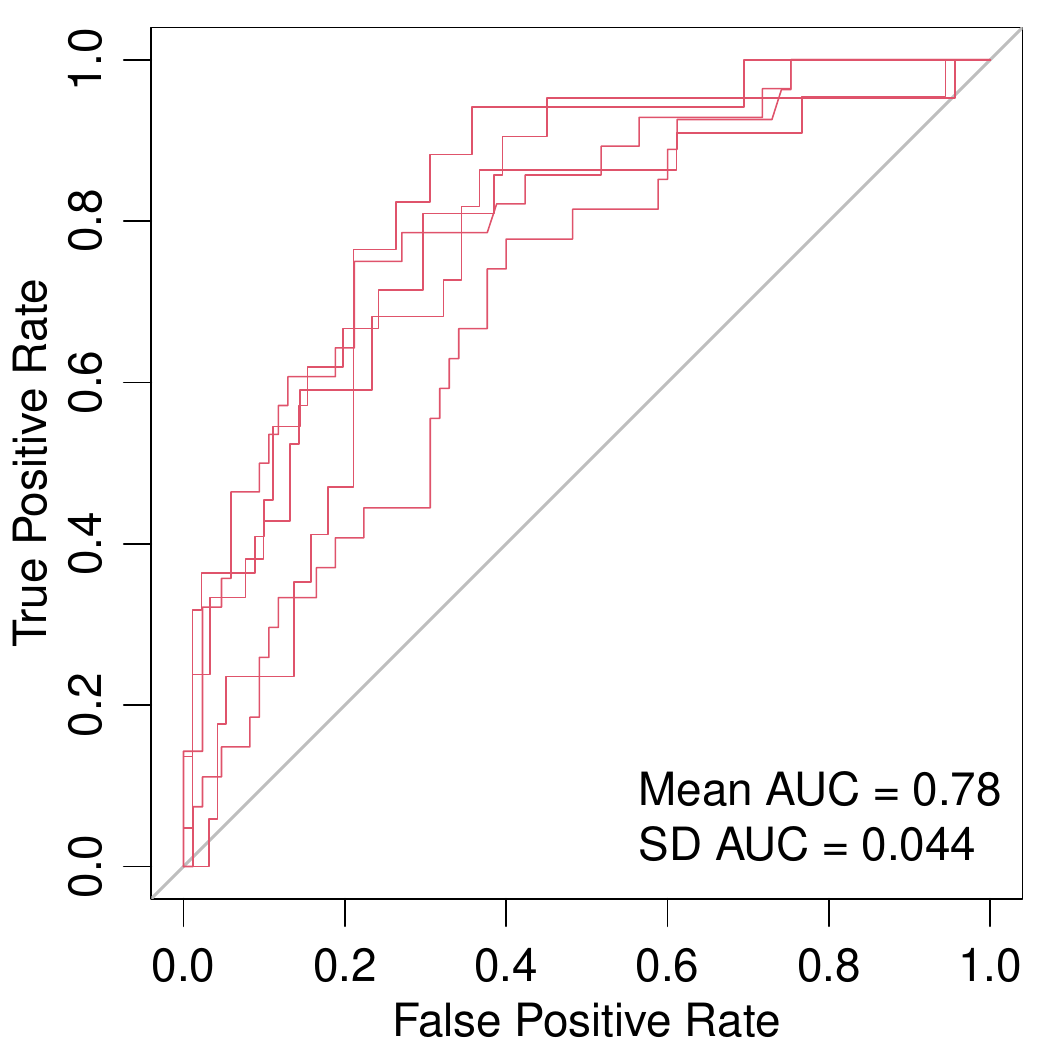}
    }
    \subfigure[Eigen model.]{
        \includegraphics[scale=0.3]{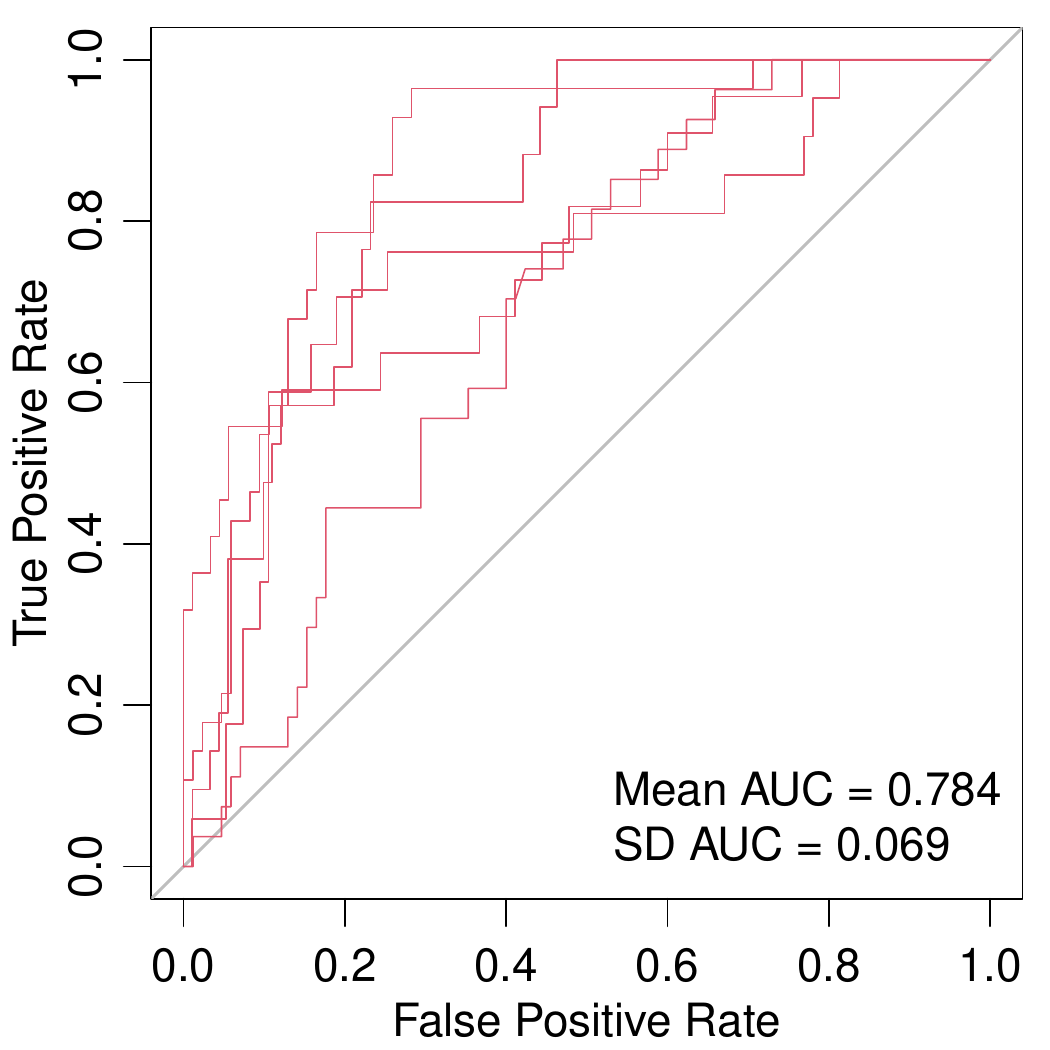}
    }
    \hspace{0em} 
    \caption{ROC curves, mean area under the curve, and standard deviation of the area under the curve from a cross-validation experiment comparing the sociality, distance, class, and eigen models using the Lazega dataset.}
    \label{fig_lazega_cv}
\end{figure}

Figure \ref{fig_lazega_cv} presents the results of the cross-validation experiment using all the models considered above on the Lazega dataset. For this dataset, we observe that the predictive performance of all models is approximately 78\%, except for the sociability model, which has a mean AUC about 6\% lower. Although its predictive capacity is slightly reduced, the results remain competitive compared to the other models. In the following section, we provide a more in-depth analysis across 20 additional datasets, where larger differences between models will become evident, and the sociability model will exhibit outstanding performance in several cases.

\begin{table}[!htb]
	\centering
	\begin{tabular}{lccccc}  
		\hline
		Acronym &  N\underline{o} actors & N\underline{o} edges & Dens. & Trans. & Assor. \\ 
		\hline
		\textsf{zach}       & 34  & 78    & 0.139 & 0.256 & -0.476 \\ 
		\textsf{bktec}      & 34  & 175   & 0.312 & 0.476 &  0.015 \\ 
		\textsf{foot}       & 35  & 118   & 0.198 & 0.329 & -0.176 \\ 
		\textsf{hitech}     & 36  & 91    & 0.144 & 0.372 & -0.087 \\ 
		\textsf{kaptail}    & 39  & 158   & 0.213 & 0.385 & -0.183 \\ 
		\textsf{bkham}      & 44  & 153   & 0.162 & 0.497 & -0.391 \\ 
		\textsf{dol}        & 62  & 159   & 0.084 & 0.309 & -0.044 \\ 
		\textsf{glossgt}    & 72  & 118   & 0.046 & 0.184 & -0.158 \\ 
		\textsf{lesmis}     & 77  & 254   & 0.087 & 0.499 & -0.165 \\ 
		\textsf{salter}     & 99  & 473   & 0.098 & 0.335 & -0.064 \\ 
		\textsf{polbooks}   & 105 & 441   & 0.081 & 0.348 & -0.128 \\ 
		\textsf{adjnoun}    & 112 & 425   & 0.068 & 0.157 & -0.129 \\ 
		\textsf{football}   & 115 & 613   & 0.094 & 0.407 &  0.162 \\ 
		\textsf{nine}       & 130 & 160   & 0.019 & 0.163 & -0.197 \\ 
		\textsf{gen}        & 158 & 408   & 0.033 & 0.078 & -0.254 \\ 
		\textsf{fblog}      & 192 & 1,431 & 0.078 & 0.386 &  0.012 \\ 
		\textsf{jazz}       & 198 & 2,742 & 0.141 & 0.520 &  0.020 \\ 
		\textsf{partner}    & 219 & 630   & 0.026 & 0.107 & -0.217 \\ 
		\textsf{indus}      & 219 & 630   & 0.026 & 0.107 & -0.217 \\ 
		\textsf{science}    & 379 & 914   & 0.013 & 0.431 & -0.082 \\ 
		\hline
	\end{tabular}
	\caption{Network datasets used for both goodness-of-fit evaluation and cross-validation experiments with the sociality, distance, class, and eigen models. ``Dens.,'' ``Trans.,'' and ``Assor.'' refer to density, transitivity, and assortativity, respectively.}\label{tab_datasets} 
\end{table}

\subsection{Model comparison using other datasets}

To compare the predictive capacity of the sociality, distance, class, and eigen models in identifying missing links, we assess their out-of-sample performance through an extensive cross-validation experiment conducted on 20 networks. These networks encompass various types, sizes, and relational structures of actors (see Table \ref{tab_datasets} for details). The datasets are freely available online from sources such as \url{http://networkrepository.com/}, \url{http://www-personal.umich.edu/~mejn/netdata/}, and \url{http://vlado.fmf.uni-lj.si/pub/networks/data/ucinet/ucidata.htm}, along with additional links provided therein. All models are fitted using the same settings as in Section 5.1 with MCMC. 

\begin{table}[!htb]
    \centering
    \begin{tabular}{lcccc|cccc}
        \hline
        \multirow{2}{*}{Network} & \multicolumn{4}{c|}{WAIC} & \multicolumn{4}{c}{AUC} \\
        \cline{2-9}
        & Soci & Dist & Class & Eigen & Soci & Dist & Class & Eigen \\
        \hline
        zach       & 386.6  & 364.6  & 330.6  & 216.3  & 0.779 & 0.714 & 0.856 & 0.858 \\
        bktec      & 636.1  & 512.4  & 621.3  & 535.6  & 0.689 & 0.777 & 0.744 & 0.715 \\
        foot       & 502.1  & 488.7  & 475.6  & 410.1  & 0.758 & 0.748 & 0.789 & 0.791 \\
        hitech     & 450.6  & 376.3  & 451.0  & 362.8  & 0.743 & 0.780 & 0.711 & 0.785 \\
        kaptail    & 680.9  & 564.6  & 643.6  & 563.6  & 0.715 & 0.788 & 0.768 & 0.726 \\
        bkham      & 458.4  & 529.3  & 454.8  & 418.8  & 0.899 & 0.824 & 0.904 & 0.859 \\
        dol        & 1061.8 & 710.9  & 912.0  & 774.2  & 0.617 & 0.811 & 0.796 & 0.744 \\
        glossgt    & 869.6  & 765.7  & 851.7  & 700.3  & 0.693 & 0.687 & 0.739 & 0.725 \\
        lesmis     & 1419.3 & 718.1  & 883.6  & 444.7  & 0.805 & 0.889 & 0.943 & 0.852 \\
        salter     & 2790.6 & 1975.5 & 2170.5 & 1969.9 & 0.731 & 0.863 & 0.869 & 0.868 \\
        polbooks   & 2867.9 & 1653.7 & 2101.7 & 1671.4 & 0.671 & 0.897 & 0.897 & 0.910 \\
        adjnoun    & 2683.5 & 2683.2 & 2660.3 & 2511.8 & 0.752 & 0.728 & 0.763 & 0.757 \\
        football   & 4150.9 & 2409.2 & 2471.7 & 2698.3 & 0.302 & 0.857 & 0.880 & 0.855 \\
        nine       & 1257.0 & 1009.0 & 1077.3 & 774.8  & 0.823 & 0.831 & 0.821 & 0.837 \\
        gen        & 2888.1 & 3250.2 & 2871.5 & 2715.2 & 0.792 & 0.675 & 0.823 & 0.713 \\
        fblog      & 9126.8 & 5251.4 & 6389.3 & 5310.1 & 0.715 & 0.926 & 0.909 & 0.762 \\
        jazz       & 13581.1& 6633.8 & 9333.5 & 6725.6 & 0.764 & 0.949 & 0.914 & 0.882 \\
        partner    & 4715.2 & 4818.8 & 4544.3 & 4447.7 & 0.814 & 0.742 & 0.845 & 0.734 \\
        indus      & 4715.2 & 4818.8 & 4544.3 & 4447.7 & 0.814 & 0.741 & 0.845 & 0.794 \\
        science    & 9390.3 & 2758.4 & 6320.8 & 873.8  & 0.644 & 0.938 & 0.922 & 0.870 \\
        \hline
    \end{tabular}
     \caption{WAIC and AUC results for the sociality, distance, class, and eigen models across different networks.}
    \label{tab_comparisson_waic_auc}
\end{table}

The results are presented in Table \ref{tab_comparisson_waic_auc}. The WAIC values assess the predictive ability of the models, with lower values indicating better performance. While the sociality model does not consistently achieve the best results, it remains competitive across many networks. The eigen model generally provides the best fit, followed by the distance and class models, which often outperform the sociality model. However, in networks such as \textit{bkham} and \textit{adjnoun}, the sociality model performs comparably to or even better than some alternatives. Although it does not minimize WAIC, its results are close to those of the best-performing models, suggesting that it effectively captures key network structures while maintaining interpretability and parsimony.
The AUC values measure predictive accuracy for missing links, where higher values indicate better performance. The sociality model remains competitive but does not consistently achieve the highest scores. The eigen and class models frequently outperform it, particularly in networks such as \textit{lesmis}, \textit{salter}, \textit{polbooks}, \textit{football}, and \textit{science}. The distance model also provides better predictions in cases such as \textit{fblog} and \textit{jazz}. However, in networks such as \textit{bkham} and \textit{nine}, the sociality model performs on par with or better than alternative models.

Although the sociality model does not achieve the lowest WAIC or the highest AUC, it remains a strong alternative due to its conceptual advantages. Unlike the distance and eigen models, which rely on latent Euclidean spaces, and the class model, which imposes categorical structures, the sociality model offers a more direct and interpretable approach to network analysis. Its consistent competitiveness across diverse datasets highlights the relevance of sociality effects, even when latent space models provide better predictive performance. While it may not always be the best option for prediction, it remains a valuable modeling framework, particularly in contexts where interpretability and the explicit representation of individual-level sociality effects are crucial.

\subsection{Simulation study}

In this section, we conduct a comprehensive simulation study to compare the sociality model's performance in terms of computation time and estimation accuracy, under MCMC and VI. To this end, we generate \( N = 100 \) random networks using the sociality model described in Section 2, with \( \mu = -2 \), which corresponds to a global connection probability of 0.022, and \( \tau^2 = 0.5 \), leading to global sociality effects centered around zero and approximately ranging from -2 to 2. These networks are generated under four different scenarios, each with a distinct sample size: 25, 50, 100, and 200. The objective is to assess how computation time and estimation accuracy vary with sample size. In each case, we compute the model's training time and the root mean squared error (RMSE) for \( \boldsymbol{\delta} = (\delta_1,\ldots,\delta_n) \). Finally, after computing these metrics for all \( N = 100 \) networks, we report the average values under each scenario (see Table \ref{tab_mcmc_vi_comparison}). All models are fitted using the same settings as in Section 5.1 with MCMC.

The results are presented in Table \ref{tab_mcmc_vi_comparison}. VI is significantly faster than MCMC while maintaining comparable accuracy. In scenario 1 (\( n = 25 \)), MCMC takes 3.26 minutes, whereas VI completes in 0.573 seconds, making it approximately 341 times faster. As the sample size increases, this difference remains substantial. For scenario 4 (\( n = 200 \)), MCMC requires 186.88 minutes, while VI completes in 65.51 seconds, maintaining a significant speed advantage. In terms of root mean squared error (RMSE), MCMC achieves slightly better precision, with values ranging from 0.692 to 0.971, compared to VI, which ranges from 0.598 to 0.884. This highlights a trade-off: while MCMC provides marginally lower estimation error, it is computationally expensive. VI, on the other hand, offers a drastic reduction in computation time with minimal loss in accuracy, making it a more practical choice for large networks.

\begin{table}[!htb]
    \centering
    \begin{tabular}{lcc|cc}
        \cline{2-5}
        \multirow{2}{*}{$n$} & \multicolumn{2}{c|}{MCMC} & \multicolumn{2}{c}{VI} \\
        \cline{2-5}
             & Time (mins) & RMSE & Time (secs) & RMSE \\
        \hline
        25   & 3.258   & 0.692 & 0.573  & 0.598 \\
        50   & 12.310  & 0.892 & 5.020  & 0.793 \\
        100  & 47.727  & 0.949 & 15.838 & 0.868 \\
        200  & 186.882 & 0.971 & 65.510 & 0.884 \\
        \hline
    \end{tabular}
    \caption{Comparison of MCMC and VI in terms of computation time and RMSE under increasing sample sizes.}
    \label{tab_mcmc_vi_comparison}
\end{table}

\section{Discussion}\label{sec_discussion}

MCMC and VI are two widely used approaches for Bayesian inference, each with distinct advantages and limitations. MCMC methods generate asymptotically samples from the posterior distribution, making them well-suited for complex models. However, they can be computationally expensive, particularly in high-dimensional spaces or large-scale networks, due to slow convergence and high autocorrelation in the samples. In contrast, VI approximates the posterior through optimization, offering greater scalability and faster computation. While VI is computationally efficient, its accuracy depends on the choice of the variational family, which can lead to biased approximations. Standard mean-field VI assumes independence among parameters, limiting its flexibility, though extensions such as stochastic variational inference (e.g., \citealt{hoffman2013stochastic}) and expectation-maximization variational inference (e.g., \citealt{vargo2015expectation}) help address these limitations by incorporating structured dependencies or stochastic updates.

Unlike models that require specifying the latent dimension, the sociality model infers node-specific effects without predefined group assignments, reducing the risk of model misspecification. Additionally, it accounts for degree heterogeneity through node-specific sociality effects, effectively separating structural dependencies from individual node attributes. By incorporating a probit-based likelihood, the sociality model facilitates efficient inference, particularly when employing variational methods.

Finally, a natural extension of the sociality model is its adaptation to directed networks, where nodes have distinct outgoing and incoming sociality parameters, resulting in an asymmetric adjacency structure. To enhance flexibility while mitigating extreme values, priors such as the Half-Cauchy distribution can be introduced for variance parameters. Further extensions include incorporating covariates, allowing sociality parameters to depend on node-specific attributes, or modeling sociality effects with a Dirichlet Process prior (e.g., \citealt{teh2010dirichlet}) to capture nonparametric degree heterogeneity. The framework can also be extended to clustering tasks by integrating hierarchical priors, enabling latent sociality parameters to define community structures. Additionally, the model can be adapted to multilayer networks, where nodes interact across different types of relationships, and dynamic networks, where latent positions and sociality effects evolve over time using either state-space models or Gaussian processes.

\section*{Statements and declarations}

The authors declare that they have no known competing financial interests or personal relationships that could have appeared to influence the work reported in this article.

During the preparation of this work the authors used ChatGPT-4-turbo in order to improve language and readability. After using this tool, the authors reviewed and edited the content as needed and take full responsibility for the content of the publication.

\bibliography{references.bib}
\bibliographystyle{apalike}

\appendix

\section{Marginal log-likelihood decomposition}\label{app_decomposition}

Since $p(\Theta\mid\mathbf{Y}) = p(\Theta,\mathbf{Y})/p(\mathbf{Y})$, it follows that
\[
\int q(\Theta; \boldsymbol{\lambda}) \log \frac{p(\mathbf{Y}, \Theta)}{q(\Theta; \boldsymbol{\lambda})} \, \textsf{d}\Theta = \int q(\Theta; \boldsymbol{\lambda}) \log \frac{p(\Theta \mid \mathbf{Y}) p(\mathbf{Y})}{q(\Theta; \boldsymbol{\lambda})} \, \textsf{d}\Theta.
\]
Expanding the logarithm,
\[
\int q(\Theta; \boldsymbol{\lambda}) \log \frac{p(\mathbf{Y}, \Theta)}{q(\Theta; \boldsymbol{\lambda})} \, \textsf{d}\Theta =
\int q(\Theta; \boldsymbol{\lambda}) \log p(\mathbf{Y}) \, \textsf{d}\Theta + \int q(\Theta; \boldsymbol{\lambda}) \log \frac{p(\Theta \mid \mathbf{Y})}{q(\Theta; \boldsymbol{\lambda})} \, \textsf{d}\Theta.
\]

Since \( \log p(\mathbf{Y}) \) is constant with respect to \( \Theta \) and \( q(\Theta; \boldsymbol{\lambda}) \) is a proper probability density, we have that:
\[
\int q(\Theta; \boldsymbol{\lambda}) \log \frac{p(\mathbf{Y}, \Theta)}{q(\Theta; \boldsymbol{\lambda})} \, \textsf{d}\Theta = \log p(\mathbf{Y}) + \int q(\Theta; \boldsymbol{\lambda}) \log \frac{p(\Theta \mid \mathbf{Y})}{q(\Theta; \boldsymbol{\lambda})} \, \textsf{d}\Theta,
\]
which is equivalent to
\[
\log p(\mathbf{Y}) = \int q(\Theta; \boldsymbol{\lambda}) \log \frac{p(\mathbf{Y}, \Theta)}{q(\Theta; \boldsymbol{\lambda})} \, \textsf{d}\Theta + \int q(\Theta; \boldsymbol{\lambda}) \log \frac{q(\Theta; \boldsymbol{\lambda})}{p(\Theta \mid \mathbf{Y})} \, \textsf{d}\Theta.
\]

\section{ELBO for the sociality model}\label{app_elbo}

The ELBO is defined as $\text{ELBO} = \textsf{P} - \textsf{Q}$,
where $\textsf{P} = \textsf{E}_{q(\Theta;\boldsymbol{\lambda})}(\log p(\mathbf{Y}, \Theta))$ is the expected log-joint probability distribution of the model under the variational distribution, and 
$\textsf{Q} = \textsf{E}_{q(\Theta;\boldsymbol{\lambda})}(\log q(\Theta;\boldsymbol{\lambda}))$ is the entropy term accounting for the complexity of the variational distribution.

On the one hand, the joint probability distribution of the model can be expressed as $\log p(\mathbf{Y}, \Theta) = \log p(\mathbf{Y} \mid \Theta) + \log p(\Theta)$. Taking the expectation under the variational distribution, we compute these components separately. First, we consider the likelihood term. The model assumes that $y_{i,j} \sim \textsf{Ber}(\Phi(\mu_{z_{i,j}}))$, with $z_{i,j} \sim \textsf{N}(\mu_{z_{i,j}}, 1)$. By integrating with respect to \( q(\Theta) \), the expected log-likelihood contributes a sum of terms of the form:
\[
-\frac{1}{2} \sum_{i<j} \left( \sigma_\mu^2 + \sigma_{\delta_i}^2 + \sigma_{\delta_j}^2 + (\mu_{z_{i,j}} - \mu_\mu - \mu_{\delta_i} - \mu_{\delta_j})^2 + \log 2\pi \right).
\]
Next, consider the prior terms. The model assumes Gaussian priors on \(\mu\) and \(\delta_i\), which contribute expectations of the form:
\[
-\frac{1}{2} \left( (\mu_\mu^2 + \sigma_\mu^2) \frac{\alpha_\sigma}{\beta_\sigma} + \psi(\alpha_\sigma) - \log \beta_\sigma + \log 2\pi \right),
\]
and
\[
-\frac{1}{2} \sum_{i} \left( (\mu_{\delta_i}^2 + \sigma_{\delta_i}^2) \frac{\alpha_\tau}{\beta_\tau} + \psi(\alpha_\tau) - \log \beta_\tau + \log 2\pi \right).
\]
Furthermore, the model assumes Inverse-Gamma priors on \(\sigma^2\) and \(\tau^2\), which contribute expectations of the form:
\[
a_\sigma \log b_\sigma - \log \Gamma(a_\sigma) - (a_\sigma + 1)(\psi(\alpha_\sigma) - \log \beta_\sigma) - b_\sigma \frac{\alpha_\sigma}{\beta_\sigma},
\]
and
\[
a_\tau \log b_\tau - \log \Gamma(a_\tau) - (a_\tau + 1)(\psi(\alpha_\tau) - \log \beta_\tau) - b_\tau \frac{\alpha_\tau}{\beta_\tau}.
\]
Summing these components, we obtain the final expression for \(\textsf{P}\).

On the other hand, the variational distribution is assumed to be factorized, allowing computation of the entropy term component-wise. The contribution from the latent probit variables \(z_{i,j}\) follows from the properties of truncated normal expectations and results in:
\[
\textsf{q}_{i,j} = 
-\frac{1}{2} \left(1 + \frac{-\mu_{z_{i,j}} \phi(-\mu_{z_{i,j}})}{1 - \Phi(-\mu_{z_{i,j}})} - \left(\frac{\phi(-\mu_{z_{i,j}})}{1 - \Phi(-\mu_{z_{i,j}})}\right)^2\right) - \log(1 - \Phi(-\mu_{z_{i,j}})),
\]
for \(y_{i,j} = 1\), and
\[
\textsf{q}_{i,j}=
-\frac{1}{2} \left(1 + \frac{-\mu_{z_{i,j}} \phi(-\mu_{z_{i,j}})}{\Phi(-\mu_{z_{i,j}})} - \left(\frac{-\phi(-\mu_{z_{i,j}})}{\Phi(-\mu_{z_{i,j}})}\right)^2\right) - \log(\Phi(-\mu_{z_{i,j}})),
\]
for \(y_{i,j} = 0\). Moreover, for the variational Gaussian distributions \(q(\mu)\) and \(q(\delta_i)\), the entropy contributions are $-\frac{1}{2} \left(\log (2\pi \sigma_\mu^2) + 1 \right)$ and $-\frac{1}{2} \sum_{i} \log \sigma_{\delta_i}^2$. For the inverse-gamma variational distributions, the entropy terms include:
\[
\alpha_\sigma \log \beta_\sigma - \log \Gamma(\alpha_\sigma) - (\alpha_\sigma + 1)(\psi(\alpha_\sigma) - \log \beta_\sigma) - \alpha_\sigma,
\]
and
\[
\alpha_\tau \log \beta_\tau - \log \Gamma(\alpha_\tau) - (\alpha_\tau + 1)(\psi(\alpha_\tau) - \log \beta_\tau) - \alpha_\tau.
\]
Summing these components, we obtain the final expression for \(\textsf{Q}\).

\section{Notation}

The cardinality of a set \(A\) is denoted by \(|A|\). If \(P\) is a logical proposition, then \(I(P) = 1\) if \(P\) is true and \(1_{\text{P}} = 0\) if \(P\) is false. The Gamma function is defined as \(\Gamma(x) = \int_0^\infty u^{x-1} e^{-u} \, \text{d}u\). Matrices and vectors whose entries consist of subscripted variables are represented using bold notation. For example, \(\boldsymbol{x} = (x_1, \dots, x_n)\) denotes an \(n \times 1\) column vector with elements \(x_1, \dots, x_n\). We use \(\boldsymbol{0}\) and \(\boldsymbol{1}\) to denote column vectors with all entries equal to 0 and 1, respectively, and \(\mathbf{I}\) to represent the identity matrix. A subscript in this context indicates the corresponding dimension; for instance, \(\mathbf{I}_n\) refers to the \(n \times n\) identity matrix. The transpose of a vector \(\boldsymbol{x}\) is denoted by \(\boldsymbol{x}^\top\), and the notation extends analogously to matrices. Additionally, for a square matrix \(\mathbf{X}\), we use \(\text{tr}(\mathbf{X})\) to denote its trace and \(\mathbf{X}^{-1}\) for its inverse. The norm of \(\boldsymbol{x}\), given by \(\sqrt{\boldsymbol{x}^\top \boldsymbol{x}}\), is denoted by \(\|\boldsymbol{x}\|\).

\begin{enumerate}

\item A random variable \(X\) has a Normal distribution with parameters \(\mu\) and \(\sigma^2\), denoted by \(X \mid \mu, \sigma^2 \sim \textsf{N}(\mu, \sigma^2)\), if its density function is
$$
p(x \mid \mu, \sigma^2) = (2\pi\sigma^2)^{-1/2} \exp\left\{ -\frac{(x - \mu)^2}{2\sigma^2} \right\},\quad x \in \mathbb{R}\,.
$$

\item A \(d \times 1\) random vector \(\boldsymbol{X} = (X_1,\ldots,X_d)\) has a multivariate Normal distribution with mean vector \(\boldsymbol{\mu}\) and covariance matrix \(\boldsymbol{\Sigma}\), denoted by \(\boldsymbol{X} \mid \boldsymbol{\mu}, \boldsymbol{\Sigma} \sim \textsf{N}_d(\boldsymbol{\mu}, \boldsymbol{\Sigma})\), if its density function is
$$
p(\boldsymbol{x} \mid \boldsymbol{\mu}, \boldsymbol{\Sigma}) = (2\pi)^{-d/2} |\boldsymbol{\Sigma}|^{-1/2} \exp\left\{ -\frac{1}{2} (\boldsymbol{x} - \boldsymbol{\mu})^\top \boldsymbol{\Sigma}^{-1} (\boldsymbol{x} - \boldsymbol{\mu}) \right\},\quad \boldsymbol{x} \in \mathbb{R}^d\,.
$$

\item A random variable \(X\) has a Truncated Normal distribution with parameters \(\mu, \sigma^2\) and truncation interval \((a, b)\), denoted by \(X \mid \mu, \sigma^2, a, b \sim \textsf{TN}(\mu, \sigma^2, a, b)\), if its density function is
$$
p(x \mid \mu, \sigma^2, a, b) = \frac{\phi\left( \frac{x - \mu}{\sigma} \right)}{\sigma \left[ \Phi\left( \frac{b - \mu}{\sigma} \right) - \Phi\left( \frac{a - \mu}{\sigma} \right) \right]},\quad a < x < b\,,
$$
where \(\phi(\cdot)\) and \(\Phi(\cdot)\) are the standard normal density and cumulative distribution functions, respectively.

\item A random variable \(X\) has a Gamma distribution with parameters \(\alpha, \beta > 0\), denoted by \(X \mid \alpha, \beta \sim \textsf{G}(\alpha, \beta)\), if its density function is
$$
p(x \mid \alpha, \beta) = \frac{\beta^\alpha}{\Gamma(\alpha)} x^{\alpha-1} \exp\left\{ -\beta x \right\},\quad x > 0\,.
$$

\item A random variable \(X\) has an Inverse Gamma distribution with parameters \(\alpha, \beta > 0\), denoted by \(X \mid \alpha, \beta \sim \textsf{IG}(\alpha, \beta)\), if its density function is
$$
p(x \mid \alpha, \beta) = \frac{\beta^\alpha}{\Gamma(\alpha)} x^{-(\alpha+1)} \exp\left\{ -\frac{\beta}{x} \right\},\quad x > 0\,.
$$

\item A random variable \(X\) has a Half-Cauchy distribution with scale parameter \(\sigma > 0\), denoted by \(X \mid \sigma \sim \textsf{HC}(\sigma)\), if its density function is
$$
p(x \mid \sigma) = \frac{2}{\pi \sigma} \left( 1 + \frac{x^2}{\sigma^2} \right)^{-1},\quad x > 0\,.
$$

\item A random variable \(X\) has a Beta distribution with parameters \(\alpha, \beta > 0\), denoted by \(X \mid \alpha, \beta \sim \textsf{Beta}(\alpha, \beta)\), if its density function is
$$
p(x \mid \alpha, \beta) = \frac{\Gamma(\alpha + \beta)}{\Gamma(\alpha)\,\Gamma(\beta)} x^{\alpha-1} (1-x)^{\beta-1},\quad 0 < x < 1\,.
$$

\item A \(K \times 1\) random vector \(\boldsymbol{X} = (X_1, \ldots, X_K)\) has a Dirichlet distribution with parameter vector \(\boldsymbol{\alpha} = (\alpha_1, \ldots, \alpha_K)\), where each \(\alpha_k > 0\), denoted by \(\boldsymbol{X} \mid \boldsymbol{\alpha} \sim \textsf{Dir}(\boldsymbol{\alpha})\), if its density function is
$$
p(\boldsymbol{x} \mid \boldsymbol{\alpha}) =
\left\{
  \begin{array}{ll}
    \frac{\Gamma\left(\sum_{k=1}^K \alpha_k\right)}{\prod_{k=1}^K \Gamma(\alpha_k)} \prod_{k=1}^K x_k^{\alpha_k - 1}, & \text{if } \sum_{k=1}^K x_k = 1; \\
    0, & \text{otherwise.}
  \end{array}
\right.
$$

\item A \(K \times 1\) random vector \(\boldsymbol{X} = (X_1, \ldots, X_K)\) has a Categorical distribution with parameter vector \(\boldsymbol{p} = (p_1, \ldots , p_K)\), where \(\sum_{k=1}^K p_k =1\), denoted by \(\boldsymbol{X} \mid \boldsymbol{p} \sim \textsf{Cat}(\boldsymbol{p})\), if its probability mass function is
$$
p(x \mid \boldsymbol{p}) =
\left\{
  \begin{array}{ll}
       \prod_{k=1}^K p_k^{\mathbb{I}(x = k)}, & \text{if } \sum_{k=1}^K x_k = 1; \\
    0, & \text{otherwise.}
  \end{array}
\right.
$$
\end{enumerate}

\end{document}